\documentclass[12pt]{article}
\usepackage{amsmath}
\begin{document}
\begin{flushright}
KANAZAWA-12-06\\
July, 2012
\end{flushright}
\vspace*{1cm}

\begin{center} 
{\Large\bf Baryon number asymmetry and dark matter 
in the neutrino mass model with an inert doublet}
\vspace*{1cm}

{\Large Shoichi Kashiwase}\footnote{e-mail:~shoichi@hep.s.kanazawa-u.ac.jp}
{\Large and Daijiro Suematsu}\footnote{e-mail:~suematsu@hep.s.kanazawa-u.ac.jp}
\vspace*{1cm}\\

{\it Institute for Theoretical Physics, Kanazawa University, 
\\ Kanazawa 920-1192, Japan}
\end{center}
\vspace*{1.5cm} 

\noindent
{\Large\bf Abstract}\\
The radiative neutrino mass model with an inert doublet scalar has been 
considered as a promising candidate which can explain neutrino masses, 
dark matter abundance and baryon number asymmetry if dark matter 
is identified with the lightest neutral component of the inert doublet. 
We reexamine these properties by imposing all the data of the neutrino 
oscillation, which are recently suggested by the reactor experiments. 
We find that the sufficient baryon number asymmetry seems not to be 
easily generated in a consistent way with all the data of the neutrino 
masses and mixing as long as the right-handed neutrinos are kept in TeV
regions. Two possible modifications of the model are examined.
\newpage
\section{Introduction}
The standard model (SM) is now considered to be extended on the basis of
several evidences clarified by recent experiments and observations, that
is, the neutrino masses and mixing \cite{nexp}, 
the existence of dark matter (DM)
\cite{uobs} and also the baryon number asymmetry in the Universe
\cite{basym}. Although there are a lot of models which are proposed to
explain these independently, it is not so easy to construct a model
which can explain all of them simultaneously without causing any other 
phenomenological problems.
If we can find such a model, it would give us crucial hints for the new
physics beyond the SM. 
The study along this line might play a crucial role for the search of 
physics beyond the SM prior to the study based on purely theoretical 
motivation such as the gauge hierarchy problem.  

The radiative neutrino mass model with an inert doublet \cite{ma} 
could be such a promising candidate. It is a very simple extension of 
the SM by an inert doublet scalar and three right-handed neutrinos only. 
An imposed $Z_2$ symmetry controls the scalar potential and 
forbids the tree-level neutrino masses since its odd parity is assigned
to these new fields and the even parity is assigned to the SM fields. 
It also guarantees the stability of the lightest 
field with its odd parity. 
Thus, the lightest neutral component of the inert doublet scalar 
\cite{idm,idm1,idm1a,inel,l5} or the lightest 
right-handed neutrino \cite{ndm,susyndm} could play a role of DM.
This feature opens a possibility for the model such that it can explain all
the above-mentioned three problems on the basis of closely related 
physics, simultaneously.
However, the model could explain only two of the three problems if
the right-handed neutrino is identified with DM \cite{ndm1}. In this case the
model is required to be extended in some way for the explanation of all
three issues \cite{ndmext}. 
On the other hand, it is noticeable that the above-mentioned 
three problems are suggested to be consistently explained by the 
new fields with TeV-scale masses as long as the lightest neutral 
component of the inert doublet scalar is identified with DM \cite{idm1}. 
This latter case seems very interesting and worthy for further 
quantitative study since the signature of the model
could be seen in various ongoing or future experiments. 
 
In this paper we reexamine this radiative neutrino mass model by 
fixing the parameters relevant to the neutrino masses and mixing 
on the basis of neutrino oscillation data including the recent results for
$\theta_{13}$ given by T2K, Double Chooz, RENO and Daya Bay \cite{t13}. 
We proceed this study by imposing the conditions required by the DM relic
abundance and its direct detection.
Based on these results, we analyze what amount of the baryon number asymmetry
can be generated via thermal leptogenesis. We show that the sufficient
baryon number asymmetry seems difficult to be generated consistently for the
parameters that are favored by the presently known phenomenological 
requirements.
 
The remaining parts of paper are organized as follows. 
In section 2, we briefly review the scalar sector of the model and then
discuss the constraints brought about by the DM relic abundance and the DM
direct detection. Next, we fix the parameters relevant to the neutrino mass
matrix to realize the neutrino oscillation data, which are
also closely related to leptogenesis.
In section 3, we apply them to the study of leptogenesis and 
estimate the baryon number asymmetry via the out-of-thermal-equilibrium 
decay of the lightest right-handed neutrino by solving the Boltzmann equations 
numerically. Conditions to generate the suitable baryon number asymmetry 
are discussed. Summary of the paper is given in section 4.

\section{DM abundance and neutrino masses}
\subsection{The model and nature of its scalar sector}
We consider the radiative neutrino mass model with an inert doublet 
scalar \cite{ma}. The model is a very simple extension of the 
standard model (SM) with three right-handed neutrinos $N_i$, and 
a scalar doublet $\eta$ which is called the inert doublet and assumed to
have no vacuum expectation value.
Although both $N_i$ and $\eta$ are supposed to have odd parity 
of an assumed $Z_2$ symmetry, all SM contents are assigned by
its even parity.
Invariant Yukawa couplings and scalar potential related to these new
fields are summarized as
\begin{eqnarray}
-{\cal L}_Y&=&h_{ij} \bar N_i\eta^\dagger\ell_{j}
+h_{ij}^\ast\bar\ell_{i}\eta N_j
+\frac{M_i}{2}\left(\bar N_iN_i^c +\bar N_i^cN_i\right), \nonumber \\
V&=&\lambda_1(\phi^\dagger\phi)^2+\lambda_2(\eta^\dagger\eta)^2
+\lambda_3(\phi^\dagger\phi)(\eta^\dagger\eta) 
+\lambda_4(\eta^\dagger\phi)(\phi^\dagger\eta) \nonumber \\ 
&+&\Big[\frac{\lambda_5}{2}(\phi^\dagger\eta)^2 + {\rm h.c.}\Big]
+m_\phi^2\phi^\dagger\phi + m_\eta^2\eta^\dagger\eta,
\label{model}
\end{eqnarray}
where $\ell_{i}$ is a left-handed lepton doublet and 
$\phi$ is an ordinary Higgs doublet. 
All the quartic coupling constants $\lambda_i$ are assumed to be real,
for simplicity. We also assume that 
neutrino Yukawa couplings $h_{ij}$ are written 
by using the basis under which both matrices for Yukawa couplings 
of charged leptons and for masses of the right-handed neutrinos are 
real and diagonal. 
These neutrino Yukawa couplings are constrained by the neutrino 
oscillation data and also the
lepton flavor-violating processes such as $\mu\rightarrow e\gamma$. 

In the following study, we assume the mass spectrum of the right-handed 
neutrinos to satisfy 
\begin{equation}
M_1 < M_2 < M_3,
\label{spectrum}
\end{equation}
and also the flavor structure of the neutrino Yukawa couplings to be
\begin{eqnarray}
&&h_{ei}=0, \quad h_{\mu i}=h_i, \quad h_{\tau i}= q_1h_i, \nonumber \\
&&h_{ej}=h_j, \quad h_{\mu j}=q_2 h_j. \quad
h_{\tau j}= -q_3 h_j,
\label{yukawa}
\end{eqnarray} 
where $q_{1,2,3}$ are real constants.
This assumption for the neutrino Yukawa couplings could reduce free 
parameters of the model substantially.
Moreover, it can cause the favorable lepton flavor mixing as found later.
We note that there remains a freedom, that is, which type structure 
represented by the suffix $i$ and $j$ in eq.~(\ref{yukawa}) should be 
assigned to each right-handed neutrino. 
In the following part, we adopt two typical cases for it as follows, 
\begin{equation}
{\rm (i)}~i=1,2,~~ j=3; \qquad {\rm (ii)}~i=1,3,~~j=2.
\label{fst}
\end{equation}

Now, we briefly review the scalar sector of the model \cite{idm,idm1}.
If we take unitary gauge and put 
$\phi^T=(0, \langle\phi\rangle +\frac{h}{\sqrt{2}})$ and 
$\eta^T=(\eta^+, \frac{1}{\sqrt 2}(\eta_R+i\eta_I))$ where 
$\langle\phi\rangle\equiv\frac{-m_\phi^2}{2\lambda_1}$, the scalar
potential $V$ in eq.~(\ref{model}) can be written as
\begin{eqnarray}
V&=&\frac{1}{2}m_h^2h^2+\frac{1}{2}M_{\eta_R}^2\eta_R^2+
\frac{1}{2}M_{\eta_I}^2\eta_I^2+ M_{\eta_c}^2\eta^+\eta^-
+\sqrt{2\lambda_1}\langle\phi\rangle h^3 \nonumber \\
&+&\frac{1}{4}\left[\sqrt{\lambda_1}h^2-\sqrt{\lambda_2}
(\eta^+\eta^-+\eta_R^2+\eta_I^2)\right]^2  \nonumber \\
&+&\frac{1}{4}h^2\left[(2\lambda_3+2\sqrt{\lambda_1\lambda_2})\eta^+\eta^-
+(2\lambda_++2\sqrt{\lambda_1\lambda_2})\eta_R^2+
(2\lambda_-+2\sqrt{\lambda_1\lambda_2})\eta_I^2\right],
\label{pot}
\end{eqnarray}
where we use the definition $\lambda_\pm=\lambda_3+\lambda_4\pm
\lambda_5$ and
\begin{equation}
m_h^2=4\lambda_1^2\langle\phi\rangle^2, 
\quad M_{\eta_c}^2=m_\eta^2+\lambda_3\langle\phi\rangle^2, \quad
M_{\eta_R}^2=m_\eta^2+\lambda_+\langle\phi\rangle^2, \quad 
M_{\eta_I}^2=m_\eta^2+\lambda_-\langle\phi\rangle^2.
\label{emass}
\end{equation}
The expression of $V$ in eq.~(\ref{pot}) shows that the assumed 
vacuum is stable for
\begin{equation}
\lambda_1,~ \lambda_2 >0, \qquad \lambda_3,~\lambda_+,~\lambda_- >
 -\sqrt{\lambda_1\lambda_2}. 
\label{cstab}
\end{equation}
We also require these quartic couplings to satisfy $|\lambda_i|<4\pi$ so
that the perturbativity of the model is guaranteed. 

Since the new doublet scalar $\eta$ is assumed to have no vacuum 
expectation value,  the $Z_2$ symmetry is kept as the unbroken symmetry 
of the model. 
Thus, the lightest field with the odd parity of this $Z_2$ is stable and then
its thermal relic behaves as DM in the Universe. 
If it is identified with $\eta_R$ here, the following condition 
should be satisfied
\begin{equation}
\lambda_4+\lambda_5<0, \qquad \lambda_5<0; \qquad M_{\eta_R} <M_1,
\label{cdm}
\end{equation}
These are easily found from eq.~(\ref{emass}).
The value of $\lambda_1$ might be estimated by using $m_h\simeq 125$~GeV,
which is suggested through the recent LHC experiments.
If we apply it to the tree-level formula in eq.~(\ref{emass}), we have 
$\lambda_1\sim 0.1$. Using this value of $\lambda_1$ and the conditions 
given in eqs.~(\ref{cstab}) and (\ref{cdm}), 
we can roughly estimate the allowed range of $\lambda_{3,4}$ as
\begin{equation}
\lambda_3>-1, \qquad 0>\lambda_4 > -4 \pi, 
\label{clambda}
\end{equation}
for the sufficiently small values of $|\lambda_5|$. The lower bound of
$\lambda_4$ is settled by the requirement for the perturbativity of the
model. 

The mass difference among the components of $\eta$ is estimated as
\begin{equation}
\frac{M_{\eta_I}-M_{\eta_R}}{M_{\eta_R}}\simeq 
\frac{|\lambda_5|\langle\phi\rangle^2}{M_{\eta_R}^2}
\equiv\frac{\delta}{M_{\eta_R}}, \qquad
\frac{M_{\eta_c}-M_{\eta_R}}{M_{\eta_R}}\simeq 
\frac{|\lambda_4+\lambda_5|\langle\phi\rangle^2}{2M_{\eta_R}^2},
\label{mdif}
\end{equation}
which could be a good approximation for the large value of $m_\eta$ 
such as $O(1)$~TeV.\footnote{Such a large value of $m_\eta$ is favored
from the analysis of the $T$ parameter for the precise measurements in
the electroweak interaction \cite{idm,idm1}. 
In that case, the model has no constraint from it.}
These formulas show that coannihilation among the components of $\eta$ 
could play an important role in the estimation of the relic 
abundance of $\eta_R$ \cite{idm1}. 

\subsection{Inert doublet dark matter}
In several articles \cite{idm}, the DM abundance is found to 
be well explained if the lightest neutral component of $\eta$ is 
identified with DM. In the high-mass $\eta$ case, in particular, 
it is suggested that the relic abundance could be a suitable value 
if one of the quartic couplings $|\lambda_i|$ in eq.~(\ref{model}) 
has magnitude of $O(1)$ \cite{idm1}.

The $\eta_R$ relic abundance is known to be estimated as \cite{relic}
\begin{equation}
\Omega_{\eta_R} h^2\simeq \frac{1.07\times 10^9{\rm GeV}^{-1}}
{J(x_F)g_\ast^{1/2} m_{\rm pl}},
\end{equation}
where the freeze-out temperature $T_F(\equiv M_{\eta_R}/x_F)$ 
and $J(x_F)$ are defined as
\begin{equation}
x_F=\ln\frac{0.038 m_{\rm pl}g_{\rm eff} M_{\eta_R}
\langle\sigma_{\rm eff}v\rangle}{(g_\ast x_F)^{1/2}}, \qquad 
J(x_F)=\int^\infty_{x_F}\frac{\langle\sigma_{\rm eff}v\rangle}{x^2}dx.
\end{equation}
The effective annihilation cross section 
$\langle\sigma_{\rm eff} v\rangle$ and the effective degrees of 
freedom $g_{\rm eff}$ are expressed 
by using the thermally averaged (co)annihilation cross section 
$\langle\sigma_{ij}v\rangle$ and the $\eta_i$ equilibrium number density 
$n_i^{\rm eq}=
\left(\frac{M_{\eta_i}T}{2\pi}\right)^{3/2}e^{-M_{\eta_i}/T}$
as\footnote{We may use the notation such as 
$(\eta_1, \eta_2, \eta_3, \eta_4)=(\eta_R, \eta_I, \eta^+, \eta^-)$ for 
convenience in the following discussion.}
\begin{equation}
\langle\sigma_{\rm eff}v\rangle=\frac{1}{g_{\rm eff}^2}
\sum_{i,j=1}^4\langle\sigma_{ij}v\rangle 
\frac{n_i^{\rm eq}}{n_1^{\rm eq}}\frac{n_j^{\rm eq}}{n_1^{\rm eq}},
\qquad
g_{\rm eff}=\sum_{i=1}^4\frac{n_i^{\rm eq}}{n_1^{\rm eq}}.
\end{equation}

The thermally averaged (co)annihilation cross section may be expanded 
by the thermally averaged relative velocity $\langle v^2\rangle$ 
of the annihilating fields as $\langle\sigma_{ij} v\rangle 
=a_{ij}+b_{ij}\langle v^2\rangle$. Since $\langle v^2\rangle \ll 1$ 
is satisfied for cold DM candidates and then $a_{ij}$ gives 
the dominant role for determining the relic abundance of $\eta_R$, 
we take account of it only neglecting the $b_{ij}$ contribution 
in this analysis.  
In the present model, the corresponding cross section is caused 
by the weak gauge interactions and also the quartic couplings $\lambda_i$. 
It is approximately calculated as \cite{idm1}
\begin{eqnarray}
a_{\rm eff}&=&\frac{(1+2c_w^4)g^4}{128\pi
 c_w^4M_{\eta_1}^2}\left(N_{11}+N_{22}+ 2N_{34}\right) \nonumber\\
&+&\frac{s_w^2g^4}{32\pi c_w^2M_{\eta_1}^2}\left(N_{13}+N_{14}+N_{23}+
N_{24}\right) \nonumber \\
&+&\frac{1}{64\pi M_{\eta_1}^2}
\left[(\lambda_+^2+\lambda_-^2+2\lambda_3^2)(N_{11}+N_{22})+
(\lambda_+-\lambda_-)^2(N_{33}+N_{44}+ N_{12})\right. \nonumber \\
&+&\left.\left\{(\lambda_+-\lambda_3)^2+
(\lambda_- -\lambda_3)^2\right\}
(N_{13}+N_{14}+N_{23}+N_{24})\right. \nonumber \\ 
&+&\left. \left\{(\lambda_+ +\lambda_-)^2
+4\lambda_3^2\right\}N_{34}\right],
\label{cross}
\end{eqnarray}
where $N_{ij}$ is defined as
\begin{equation}
N_{ij}\equiv\frac{1}{g_{\rm eff}^2}
\frac{n_i^{\rm eq}}{n_1^{\rm eq}}\frac{n_j^{\rm eq}}
{n_1^{\rm eq}}
=\frac{1}{g_{\rm eff}^2}
\left(\frac{M_{\eta_i}M_{\eta_j}}{M_{\eta_1}^2}\right)^{3/2}
\exp\left[-\frac{M_{\eta_i}+M_{\eta_j}-2M_{\eta_1}}{T}\right].
\label{eqfactor}
\end{equation}
Using these formulas, we examine the condition on 
the relevant parameters of the model to realize the relic
abundance $\Omega_{\eta_R}h^2=0.11$ which is required 
from the WMAP data \cite{uobs}.
In Fig.~1, we plot $\Omega_{\eta_R}h^2$ as a function of $\lambda_4$ 
for $\lambda_5=-10^{-5}$ and some typical values of $\lambda_3$.
We should note that the allowed region of $\lambda_4$ is 
restricted by eq.~(\ref{clambda}).
This figure shows that the required value for $\Omega_{\eta_R}h^2$ could 
be obtained for a wide range value of $m_\eta$ 
if $|\lambda_3+\lambda_4|$ has a value of $O(1)$.  
Since we consider the high-mass region such as 
$m_{\eta} \gg \langle\phi\rangle$,
the coannihilation among the components of $\eta$ could be effective to
reduce the relic abundance of $\eta_R$.

\input epsf
\begin{figure}[t]
\begin{center}
\epsfxsize=7.5cm
\leavevmode
\epsfbox{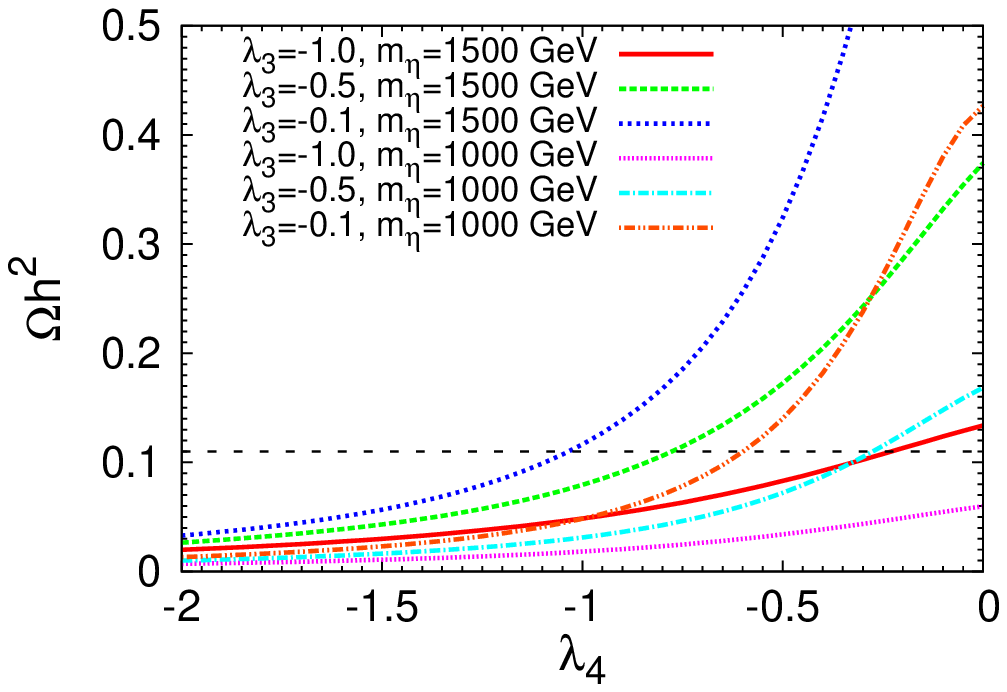}
\hspace*{5mm}
\epsfxsize=7.5cm
\leavevmode
\epsfbox{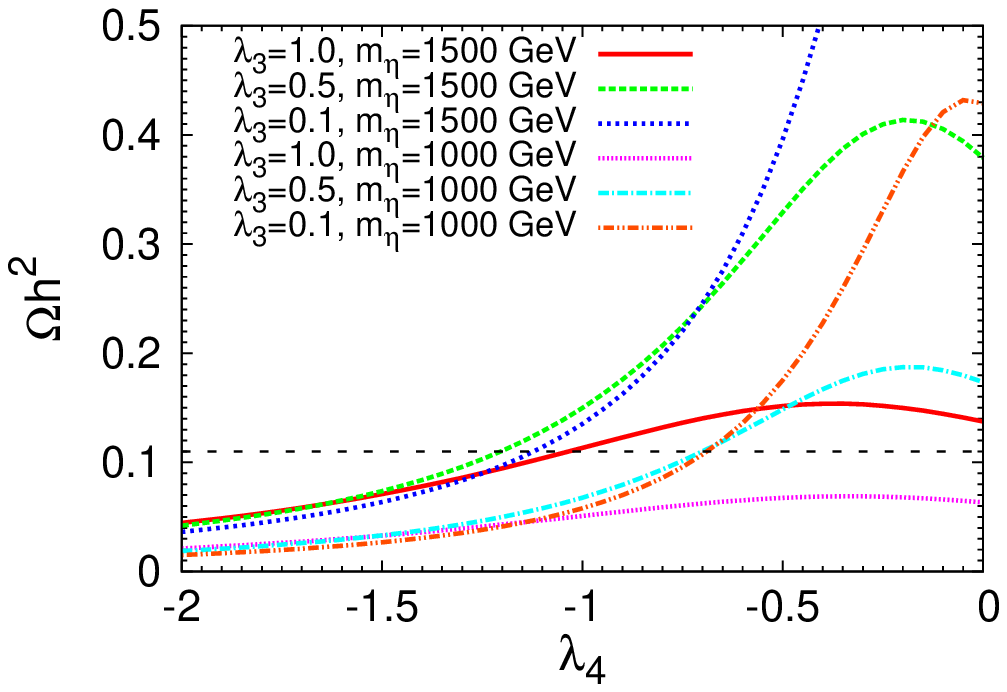}
\end{center}
\vspace*{-3mm}

{\footnotesize {\bf Fig.~1}~~$\Omega_{\eta_R}h^2$ as a function of 
$\lambda_4$ for the negative value of $\lambda_3$ such as $-1,~-0.5,~-0.1$ 
in the left panel and for the positive value of $\lambda_3$ such as 
$1,~0.5,~0.1$ in the right panel.  
The value of $m_\eta$ is fixed to 1000 and 1500~GeV in both cases.}  
\end{figure}

The above analysis shows that the DM relic abundance gives only a 
weak condition on some of the quartic couplings but no conditions 
on the neutrino Yukawa couplings. 
This is completely different from the case in which the lightest $N_i$
is identified with DM \cite{ndm,susyndm,ndm1}. 
On the other hand, the direct search of DM could give a severe constraint 
on the value of $\lambda_5$, which plays a crucial role in this radiative 
neutrino mass generation. 
Elastic scatterings between $\eta_R$ and nucleus could be mediated 
by the Higgs exchange at tree level and also by the gauge boson
exchange at one-loop level. 
However, their effects are much smaller than
the present upper bounds of the sensitivity for the direct detection.
Thus, we can neglect their effects in this discussion.
On the other hand, inelastic scattering of $\eta_R$ with nucleus 
 mediated by the $Z^0$ exchange could bring about an important 
effect to the direct search experiments \cite{inel,l5}, 
since the masses of $\eta_R$ and $\eta_I$ are almost 
degenerate for small values of $|\lambda_5|$ as found from
eq.~(\ref{mdif}). 

If we note that the interaction of $\eta_{R}$ 
relevant to this process is given by
\begin{equation}
{\cal L}=g\eta_R\partial_\mu\eta_IZ^\mu - g\eta_I\partial_\mu\eta_RZ^\mu,
\end{equation}
it is found that the inelastic nucleus-DM scattering can occur for the
DM with velocity larger than a minimum value given by \cite{inelvel}
\begin{equation}
v_{\rm min}=\frac{1}{\sqrt{2m_NE_R}}\left(\frac{m_NE_R}{\mu_N}
+\delta\right),
\end{equation}
where $\delta$ is the mass difference between $\eta_R$ and $\eta_I$ 
defined in eq.(\ref{mdif}). $E_R$ is the nucleus recoil
energy, and $m_N$ and $\mu_N$ are the mass of the target nucleus and the
reduced mass of the nucleus-DM system.
Thus, the mass difference $\delta$ is constrained by the fact that 
no DM signal is found in the direct 
DM search yet \cite{direct1,direct2}.\footnote{The DAMA data have 
been suggested
to be explained by the DM inelastic scattering \cite{inel,l5}. 
However, we do not consider it here. }
This condition might be estimated as $\delta~{^>_\sim}150$~keV \cite{l5}. 
Since $\delta$ is related to $\lambda_5$ through eq.(\ref{mdif}) , 
this constrains the allowed value of $|\lambda_5|$ 
such as\footnote{We should note that the bound of $\delta$ largely 
depends on the DM velocity in the neighborhood of the Earth. If we take 
$\delta~{^>_\sim}1$~MeV, this inelastic scattering effect can
be completely neglected even for the maximally estimated DM velocity.
In that case, the lower bound of $|\lambda_5|$ becomes one order of
magnitude larger.} 
\begin{equation}
|\lambda_5|\simeq \frac{M_{\eta_1}\delta}{\langle\phi\rangle^2} 
~{^>_\sim}~5.0\times 10^{-6}
\left(\frac{M_{\eta_1}}{1~{\rm TeV}}\right)
\left(\frac{\delta}{150~{\rm keV}}\right).
\label{direct}
\end{equation}
We take account of this constraint in the following analysis of 
the neutrino masses and the baryon number asymmetry.

\subsection{Neutrino masses and mixing}
Neutrino masses are generated through one-loop diagrams with 
the contribution of new $Z_2$ odd fields. They can be expressed as 
\cite{ma,ndm}
\begin{equation}
{\cal M}^\nu_{ij}=\sum_{k=1}^3h_{ik}h_{jk}
\left[\frac{\lambda_5\langle\phi\rangle^2}
{8\pi^2M_k}\frac{M_k^2}{M_\eta^2-M_k^2}
\left(1+\frac{M_k^2}{M_\eta^2-M_k^2}\ln\frac{M_k^2}{M_\eta^2}\right)\right]
\equiv \sum_{k=1}^3h_{ik}h_{jk}\Lambda_k,
\label{nmass}
\end{equation}
where $M_\eta^2=m_\eta^2+(\lambda_3+\lambda_4)\langle\phi\rangle^2$.
Since we consider the high-mass region such as 
$m_\eta \gg\langle\phi\rangle$, the mass difference among $\eta_i$ 
caused by nonzero $\lambda_4$ is negligible in the neutrino mass 
analysis. Thus, we treat their masses as $M_\eta$.
Both neutrino masses and mixing are determined by the couplings 
$\lambda_5$ and $h_{ik}$, the right-handed neutrino masses $M_i$'s and
the inert doublet mass $M_\eta$.
Here we note that the neutrino Yukawa couplings could take rather large values
even for the light right-handed neutrinos with masses of $O(1)$~TeV as long
as $|\lambda_5|$ takes a small value in the range given by 
eq.~(\ref{direct}). This freedom is crucial when 
we consider leptogenesis in this model as seen in the next section.

Now we have a lot of information on the feature of lepton flavor 
mixing on the basis of the neutrino oscillation data including the
recent results for $\theta_{13}$ \cite{prg}. 
We can use it to restrict the neutrino Yukawa couplings.
The flavor structure of the neutrino Yukawa couplings assumed in
eq.~(\ref{yukawa}) makes the neutrino mass matrix take a simple form such as
\begin{equation}
{\cal M}^\nu=
\left(
\begin{array}{ccc}
0 & 0 & 0\\ 0 & 1 & q_1 \\ 0 & q_1 & q_1^2 \\ 
\end{array}\right)(h_1^2\Lambda_1+ h_i^2\Lambda_i)+
\left(
\begin{array}{ccc}
1 & q_2 & -q_3\\ q_2 & q_2^2 & -q_2q_3 \\ -q_3 & -q_2q_3 & q_3^2 \\ 
\end{array}\right)h_j^2\Lambda_j,
\label{nmass2}
\end{equation} 
where $i,j$ should be understood to stand for (i) $i=2,~j=3$ 
and (ii) $i=3,~j=2$ following eq.~(\ref{fst}). 
If we put $q_{1,2,3}=1$ in both cases, the PMNS mixing matrix is 
easily found to have a tribimaximal form 
\begin{equation}
U_{PMNS}=\left(\begin{array}{ccc}
\frac{2}{\sqrt 6} & \frac{1}{\sqrt 3} & 0\\
 \frac{-1}{\sqrt 6} & \frac{1}{\sqrt 3} & \frac{1}{\sqrt 2}\\
\frac{1}{\sqrt 6} & \frac{-1}{\sqrt 3} & \frac{1}{\sqrt 2}\\
\end{array}\right)
\left(\begin{array}{ccc}
1 &0 & 0\\
0 & e^{i\alpha_1} & 0 \\
0 & 0 & e^{i\alpha_2} \\
\end{array}\right),
\label{mns} 
\end{equation}
where Majorana phases $\alpha_{1,2}$ are determined by the phases 
$h_i$ and $\lambda_5$. If we put $\varphi_i={\rm arg}(h_i)$ and 
$\varphi_{\lambda_5}={\rm arg}(\lambda_5)$, they are expressed as
\begin{equation}
\alpha_1=\varphi_3 +\frac{\varphi_{\lambda_5}}{2}, \quad
\alpha_2=\frac{1}{2}\tan^{-1}
\left(\frac{|h_1|^2\Lambda_1\sin(2\varphi_1+\varphi_{\lambda_5})+
|h_2|^2\Lambda_2\sin(2\varphi_2+\varphi_{\lambda_5})}
{|h_1|^2\Lambda_1\cos(2\varphi_1+\varphi_{\lambda_5})
+|h_2|^2\Lambda_2\cos(2\varphi_2+\varphi_{\lambda_5})}\right).
\end{equation}
In this case, one of mass eigenvalues is zero. Thus,
if $|h_1|$ is assumed to take a sufficiently small 
value compared with others, we find that the mass eigenvalues should satisfy
\begin{equation}
|h_i|^2\Lambda_i \simeq 
\frac{\sqrt{\Delta m_{\rm atm}^2}}{2}, \qquad
|h_j|^2\Lambda_j\simeq \frac{\sqrt{\Delta m_{\rm sol}^2}}{3},
\label{c-oscil}
\end{equation}
where $\Delta m^2_{\rm atm}$ and $\Delta m^2_{\rm sol}$ stand for 
the squared mass differences required by the neutrino oscillation analysis
for both atmospheric and solar neutrinos \cite{nexp,prg}.\footnote{We
confine our study to the normal hierarchy here.}

\begin{figure}[t]
\begin{center}
\epsfxsize=7.5cm
\leavevmode
\epsfbox{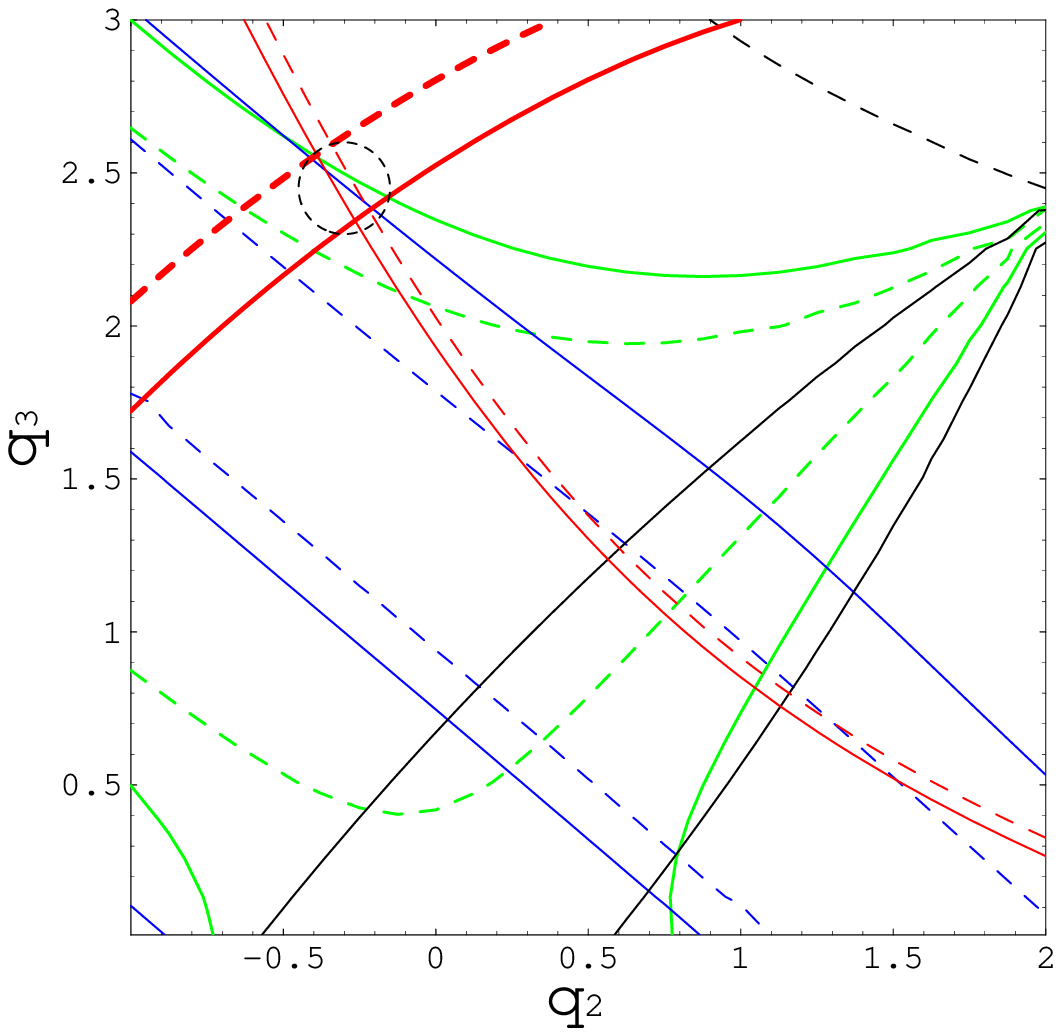}
\hspace*{2mm}
\epsfxsize=7.5cm
\leavevmode
\epsfbox{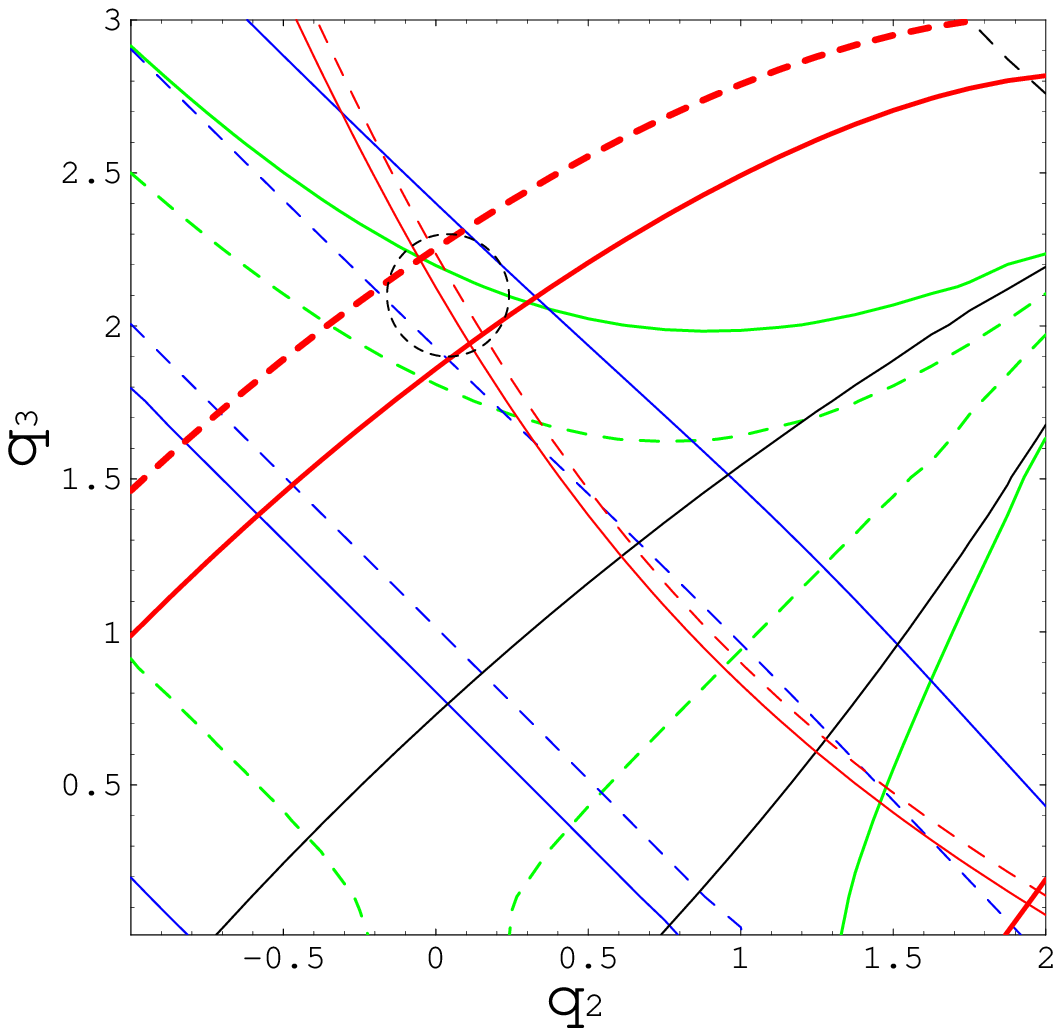}
\end{center}
\vspace*{-3mm}

{\footnotesize {\bf Fig.~2}~~Regions in the $(q_2,q_3)$ plane
allowed by the neutrino oscillation data for $q_1=0.85$ (left panel) and
 $q_1=1$ (right panel), which are contained in the circle drawn by the
 dotted line. Relevant parameters are fixed to the ones 
shown in Table 1. Each contour in both panels represents 
2$\sigma$ boundary values of neutrino oscillation parameters 
$\Delta m^2_{32}$ (thick red solid
 and dashed lines), $|\Delta m_{12}^2|$ (thin red solid and dashed lines), 
$\sin^22\theta_{23}$ (greed solid and dashed lines), 
$\sin^22\theta_{12}$ (blue solid and dashed lines) which are 
given in Ref.\cite{bestf}. The 90\% CL value of $\sin^22\theta_{13}$ given 
in Ref.\cite{t13} is also plotted as a reference (black solid 
and dashed lines).}   
\end{figure}

Using the formulas (\ref{mns}) and (\ref{c-oscil}), we can examine whether 
parameters obtained in the previous part could be consistent with 
the neutrino oscillation data. This gives 
a useful starting point for the analysis. 
However, if we take account of the fact that $\theta_{13}$ is found to 
have a nonzero value now, we cannot use them in the analysis directly. 
Here, we numerically diagonalize the mass matrix (\ref{nmass2}) and
impose both the neutrino oscillation data and the constraint 
on $|\lambda_5|$ given in eq.~(\ref{direct}) to restrict the 
neutrino Yukawa couplings. 
By fixing $q_1$ to typical values, in Fig.~2, we plot contours 
in the $(q_2, q_3)$ plane which correspond to 2$\sigma$ bounds 
of the neutrino oscillation parameters given in \cite{bestf}.  
This figure shows the mass matrix (\ref{nmass2}) can explain
all the neutrino oscillation data consistently at the regions in the
$(q_2, q_3)$ plane, which are the region including $(-0.27,2.4)$ in the
left panel ($q_1=0.85$) and the region including $(0.05,2.1)$ in the
right panel ($q_1=1$).
These regions in two panels are obtained for each set of parameters
listed in Table 1. 
We also give the predicted values for $\sin^22\theta_{13}$ in each case of
the same Table.
These examples show that neutrino Yukawa couplings of $O(10^{-3})$ can 
explain the neutrino oscillation data for $|\lambda_5|=O(10^{-5})$ and
the right-handed neutrinos with the mass of $O(1)$ TeV.

Lepton flavor-violating processes such as $\mu\rightarrow e\gamma$ 
are also induced through one-loop diagrams which have $\eta$ and 
$N_i$ in the internal lines \cite{ndm}. 
Their present experimental bounds could impose severe constraints 
on the model depending on the values of neutrino Yukawa couplings $h_i$.
However, since the small $|h_i|$ of $O(10^{-3})$ can realize the 
appropriate values for neutrino masses even for the TeV-scale values 
of $M_i$ and $M_\eta$ as discussed above, the new contributions to 
the lepton flavor-violating processes are sufficiently suppressed 
such as ${\rm Br}(\mu\rightarrow e\gamma)=O(10^{-18})$ and 
${\rm Br}(\tau\rightarrow \mu\gamma)=O(10^{-14})$.
These values show that the lepton flavor-violating processes
bring about no substantial constraints on the model. 
We should note that the freedom of $\lambda_5$ in this mass generation 
scheme makes it possible.
     
\begin{figure}[t]
\begin{center}
\begin{tabular}{c|cccccccccc}\hline
& $q_1$ &$M_\eta$ & $M_1$ & $M_2$ & $M_3$ & 
$10^3|h_2|$ & $10^3|h_3|$ & $\sin^22\theta_{13}$ & 
max$|\varepsilon_{2,3}|$ & $Y_B$\\ \hline
(ia)  &0.85 & 1 &  2  & 6   &  10   & 3.41   & 1.50 & 0.085 &
$1.1\cdot 10^{-7}$ &$2.7\cdot 10^{-12}$\\
(ib)  &0.85 & 1 &  2  & 20  & 200  & 4.62   & 4.16 & 0.085 &
$6.1\cdot 10^{-8}$ &$1.6\cdot 10^{-12}$ \\
(ic)  & 1 & 1 &  2  & 6   &  10   & 3.41   & 1.50  & 0.053 &
$9.8\cdot 10^{-8}$ & $2.8\cdot 10^{-12}$ \\
(iia) &0.85 & 1 &  2  & 6   &  10   & 1.34  & 3.81 & 0.085 &
$1.7\cdot 10^{-8}$ & $2.7 \cdot 10^{-13}$ \\
(iib) &0.85 & 1 &  2  & 20  & 200  & 1.82  & 10.6 & 0.085 &
$9.5\cdot 10^{-9}$ & $8.3 \cdot 10^{-13}$ \\ \hline
\end{tabular}
\end{center}
\vspace*{3mm}

{\footnotesize{\bf Table 1}~ The predicted value of $\sin^22\theta_{13}$
and $Y_B$ for the model parameters that can
 satisfy the neutrino oscillation data. Cases (i) and (ii) correspond 
to the ones defined in eq.~(\ref{fst}).
In all cases, $|\lambda_5|$ and $|h_1|$ are 
fixed to $10^{-5}$ and $3\cdot 10^{-8}$, respectively. The value of 
$\sin^22\theta_{13}$ is evaluated at $(q_2,q_3)=(-0.27,2.4)$ for 
$q_1=0.85$ and $(0.05,2.1)$ for $q_1=1$, where all other neutrino
 oscillation data are satisfied. A TeV unit is used as the mass scale.}
\end{figure}

\section{Baryon number asymmetry}
\subsection{Leptogenesis via the decay of the TeV-scale right-handed neutrino}
We consider the thermal leptogenesis \cite{fy} 
in this model with the mass spectrum 
given in eqs.~(\ref{spectrum}) and (\ref{cdm}).
In this case, the lepton number asymmetry is expected to be 
generated through the out-of-thermal-equilibrium decay of 
the right-handed neutrino $N_1$.
The dominant contribution to the $CP$ asymmetry $\varepsilon$ in 
this decay is brought about 
by the interference between the tree diagram and the one-loop vertex 
diagram as usual. However, we should note that 
the $\eta$ mass is not negligible compared with the one of $N_1$ in this
model. Taking account of this feature, $\varepsilon$ can be calculated 
as \cite{epsilon}
\begin{eqnarray}
\varepsilon&=&\frac{1}{16\pi
\left[\frac{3}{4}+\frac{1}{4}\left(1-\frac{M_\eta^2}{M_1^2}\right)^2\right]}
\sum_{i=2,3}\frac{{\rm Im}
\left[\left(\sum_{k=e,\mu,\tau}h_{k1}h^\ast_{ki}\right)^2\right]}
{\sum_{k=e,\mu,\tau}h_{k1}h_{k1}^\ast}
G\left(\frac{M_i^2}{M_1^2},\frac{M_\eta^2}{M_1^2}\right)\nonumber \\
&\equiv &\varepsilon_2\sin 2(\varphi_1-\varphi_2)+
\varepsilon_3\sin 2(\varphi_1-\varphi_3),
\label{cp}
\end{eqnarray}
where $G(x,y)$ is defined by
\begin{equation}
G(x,y)=\frac{5}{4}F(x,0)+\frac{1}{4}F(x,y)
+\frac{1}{4}(1-y)^2\left[F(x,0)+F(x,y)\right], 
\end{equation}
and $F(x,y)$ is represented as
\begin{equation}
F(x,y)=\sqrt{x}\left[1-y-(1+x)\ln\left(\frac{1-y+x}{x}\right)\right].
\end{equation}
If we use the flavor structure of neutrino Yukawa
couplings (i) and (ii) given in eq.~(\ref{fst}), 
$\varepsilon_{2,3}$ are expressed for each case as
\begin{eqnarray}
&{\rm (i)}& \varepsilon_2=C(1+q_1^2)|h_2|^2 
G\left(\frac{M_2^2}{M_1^2},\frac{M_\eta^2}{M_1^2}\right), \ 
\varepsilon_3=\frac{C(q_2-q_1q_3)^2|h_3|^2}{1+q_1^2}
G\left(\frac{M_3^2}{M_1^2},\frac{M_\eta^2}{M_1^2}\right), \nonumber \\
&{\rm (ii)}& \varepsilon_2=\frac{C(q_2-q_1q_3)^2|h_2|^2}{1+q_1^2} 
G\left(\frac{M_2^2}{M_1^2},\frac{M_\eta^2}{M_1^2}\right), \ 
\varepsilon_3=C(1+q_1^2)|h_3|^2
G\left(\frac{M_3^2}{M_1^2},\frac{M_\eta^2}{M_1^2}\right),
\end{eqnarray}
where $C^{-1}=16\pi\left[\frac{3}{4}+\frac{1}{4}
\left(1-\frac{M_\eta^2}{M_1^2}\right)^2\right]$.

The decay of $N_1$ should be out of equilibrium
so that the lepton number asymmetry is generated through it. 
If we express the Hubble parameter and the decay width of $N_1$
by $H$ and $\Gamma_{N_1}^D$ respectively,
this condition is given as $H>\Gamma_{N_1}^D$ at $T\sim M_1$ 
where the lepton number asymmetry is considered to be
dominantly generated.
Since $\Gamma_{N_1}^D$ is expressed as 
$\Gamma_{N_1}^D=\frac{|h_1|^2}{8\pi}(1+q_1)^2
M_1\left(1-\frac{M_\eta^2}{M_1^2}\right)^2$, 
we find that the Yukawa coupling $|h_1|$ should be 
sufficiently small such as  
\begin{equation}
|h_1|< 2\times 10^{-8}(1+q_1^2)^{-1/2}
\left(\frac{M_1}{1~{\rm TeV}}\right)^{1/2}.
\label{h1}
\end{equation}
We note that this constraint could be weaker since both $M_\eta$ and the 
Boltzmann suppression factor are neglected in this estimation. 
As found in the numerical calculation, $|h_1|$ can be somewhat larger 
than this bound. 

The generated lepton number asymmetry could be washed out by both the 
lepton number violating 2-2 scattering such as 
$\eta\eta\rightarrow\ell_\alpha\ell_\beta$ and
$\eta\ell_\alpha\rightarrow\eta^\dagger\bar\ell_\beta$ 
and also the inverse decay of
$N_1$. If the relevant Yukawa couplings are much smaller than $O(1)$,
these processes are expected to decouple before the temperature $T$ 
of the thermal plasma decreases to $T\sim M_1$. 
In order to study this quantitatively, 
we numerically solve the coupled Boltzmann equations for the number 
density of $N_1$ and the lepton number asymmetry which are expressed 
by $n_{N_1}$ and $n_L$ here, respectively.
The Boltzmann equations for these quantities are written as \cite{kt}
\begin{eqnarray}
&&\frac{dY_{N_1}}{dz}=-\frac{z}{sH(M_1)}
\left(\frac{Y_{N_1}}{Y_{N_1}^{\rm eq}}-1\right)\left\{
\gamma_D^{N_1}+\sum_{i=2,3}\left(\gamma_{{N_1}{N_i}}^{(2)}+
\gamma_{{N_1}{N_i}}^{(3)}\right)\right\}, \nonumber \\
&&\frac{dY_L}{dz}=\frac{z}{sH(M_1)}\left\{
\varepsilon\left(\frac{Y_{N_1}}{Y_{N_1}^{\rm eq}}-1\right)\gamma_D^{N_1}
-\frac{2Y_L}{Y_\ell^{\rm eq}}\left(\gamma_N^{(2)}
+\gamma_N^{(13)}\right)\right\}, 
\label{bqn}
\end{eqnarray}
where $z=\frac{M_1}{T}$ and 
$H(M_1)=1.66g_\ast^{1/2}\frac{M_1^2}{m_{\rm pl}}$.
$Y_{N_1}$ and $Y_L$ are defined as $Y_{N_1}=\frac{n_{N_1}}{s}$ 
and $Y_L=\frac{n_L}{s}$ 
by using the entropy density $s$. 
Their equilibrium values are expressed as 
$Y_{N_1}^{\rm eq}(z)=\frac{45}{2\pi^4g_\ast}z^2K_2(z)$
and $Y_\ell^{\rm eq}=\frac{45}{\pi^4g_\ast}$, where $g_\ast$ is the
number of relativistic degrees of freedom and $K_2(z)$ is the
modified Bessel function of the second kind.
In these equations we omit terms whose contributions are considered to be 
negligible compared with others.
The formulas of the relevant reaction density $\gamma$ contained in 
these equations are given in the Appendix.
If we use the relation $B= \frac{8}{23}(B-L)$ which is derived 
with the chemical equilibrium condition in this model,
the baryon number asymmetry $Y_B(=\frac{n_B}{s})$ in the present
Universe is found to be estimated as
\begin{equation}
Y_B=-\frac{8}{23}Y_L(z_{\rm EW})
\label{baryon}
\end{equation}
by using the solution $Y_L(z)$ of the coupled equations in eq.~(\ref{bqn}).
Here $z_{\rm EW}$ is related to the sphaleron decoupling 
temperature $T_{\rm EW}$ as $z_{\rm EW}=\frac{M_1}{T_{\rm EW}}$.  
 
\begin{figure}[t]
\begin{center}
\epsfxsize=4.8cm
\leavevmode
\epsfbox{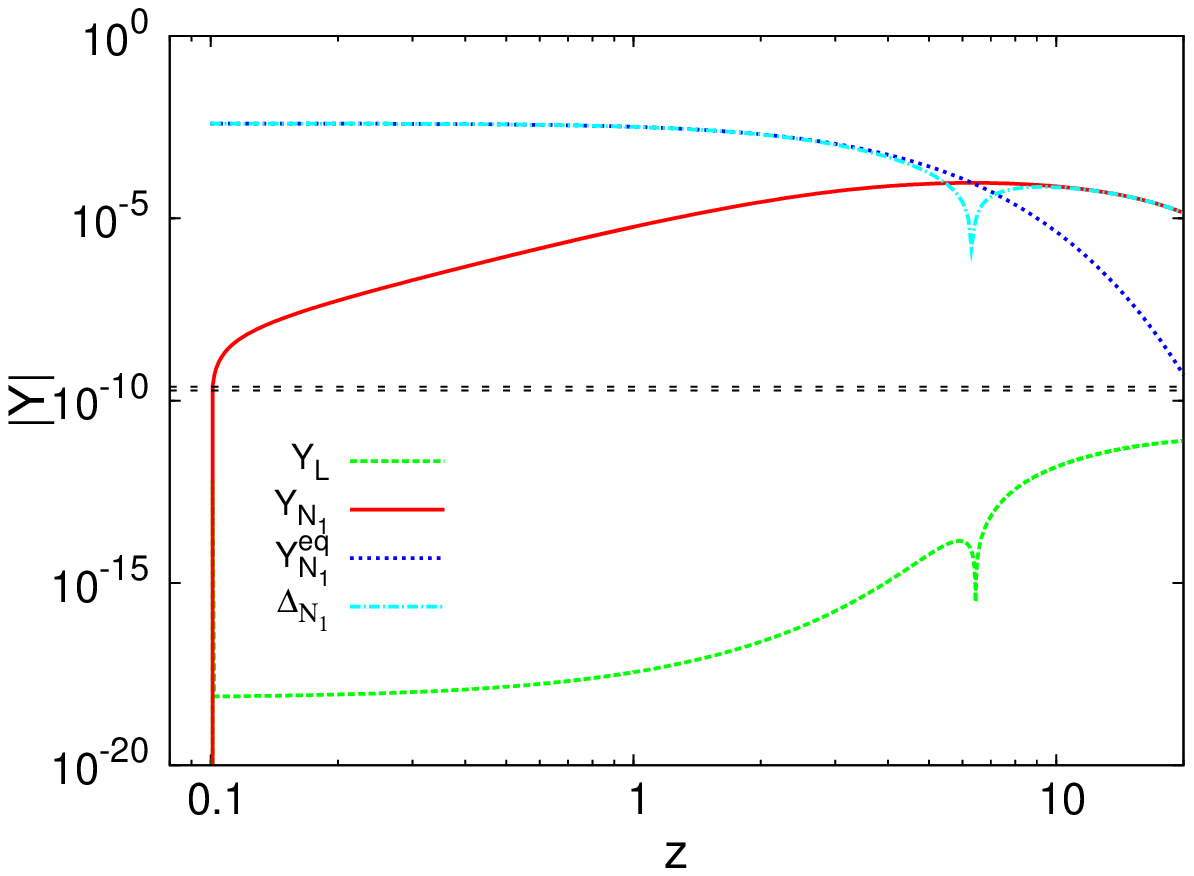}
\hspace*{3.7mm}
\epsfxsize=4.8cm
\leavevmode
\epsfbox{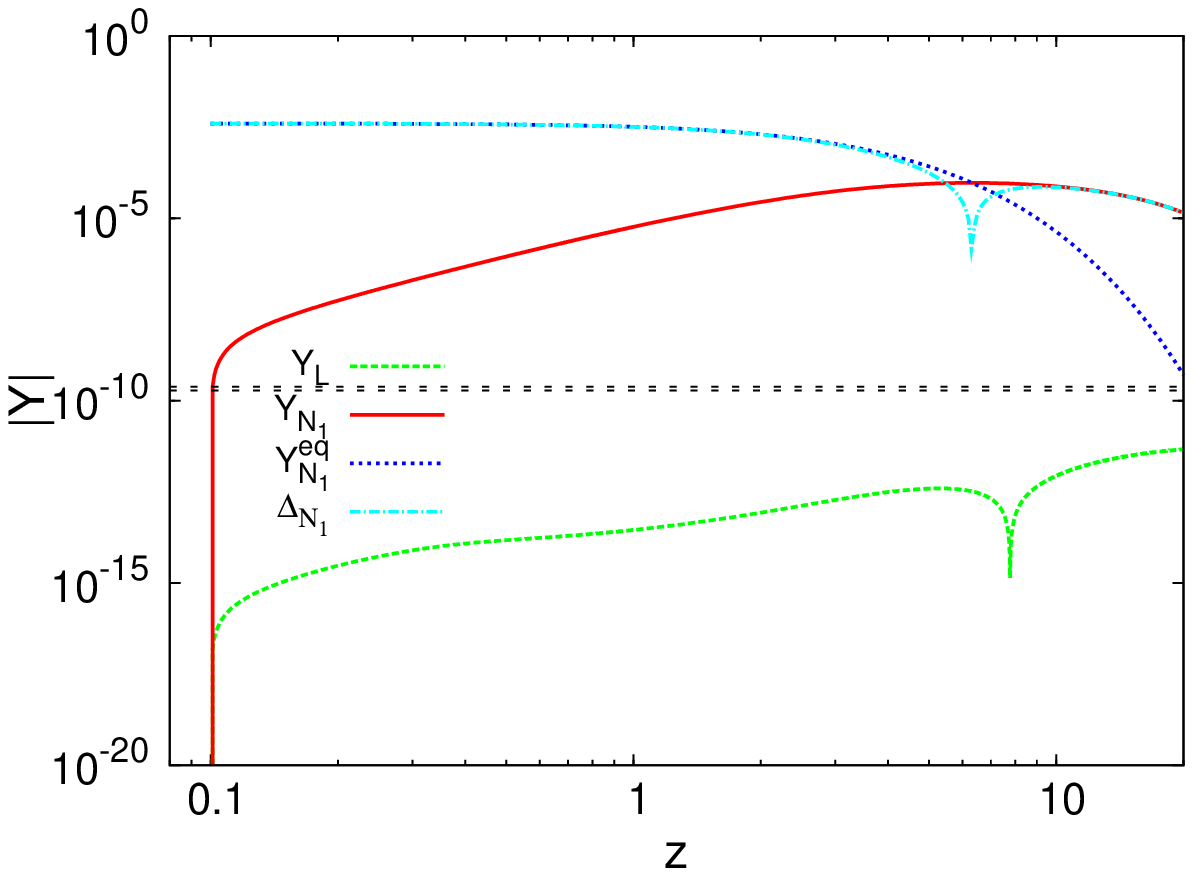}
\hspace*{3.7mm}
\epsfxsize=4.8cm
\leavevmode
\epsfbox{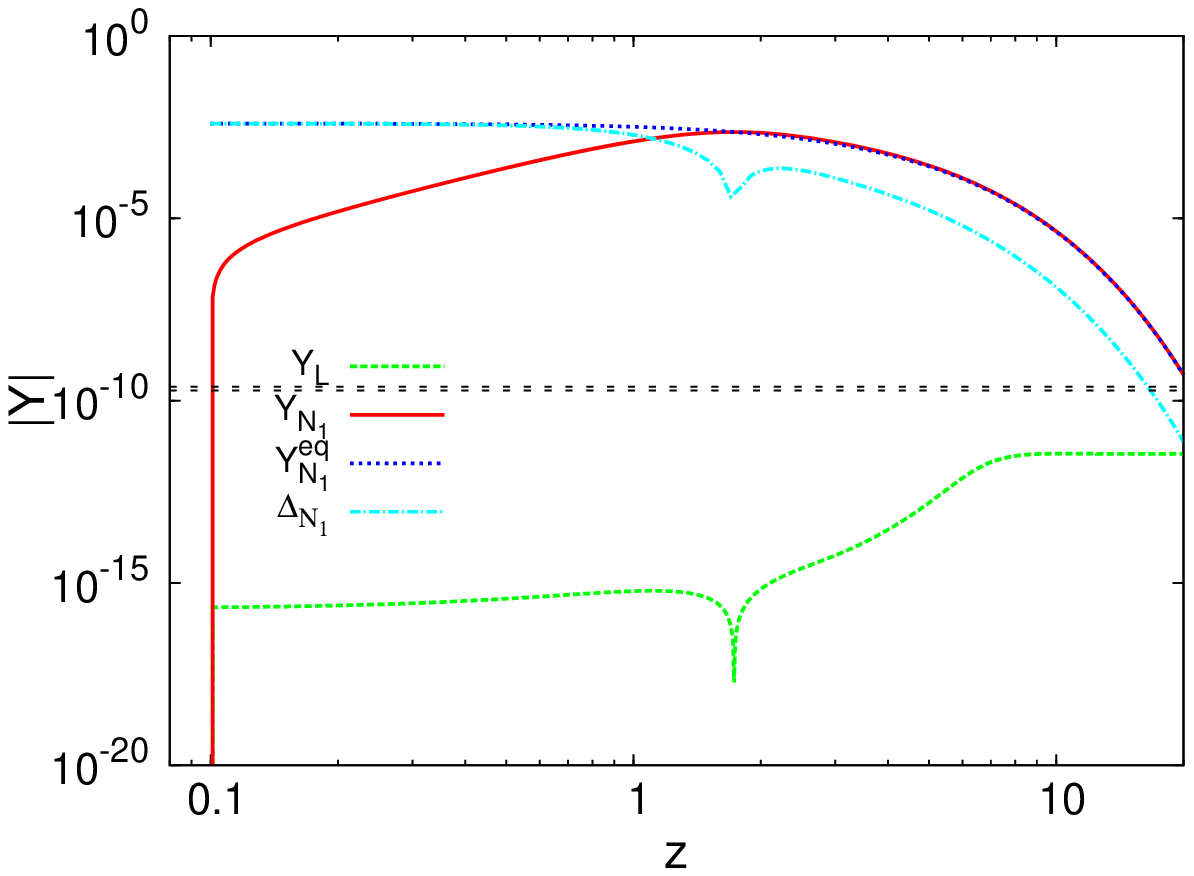} \\
\epsfxsize=4.7cm
\leavevmode
\epsfbox{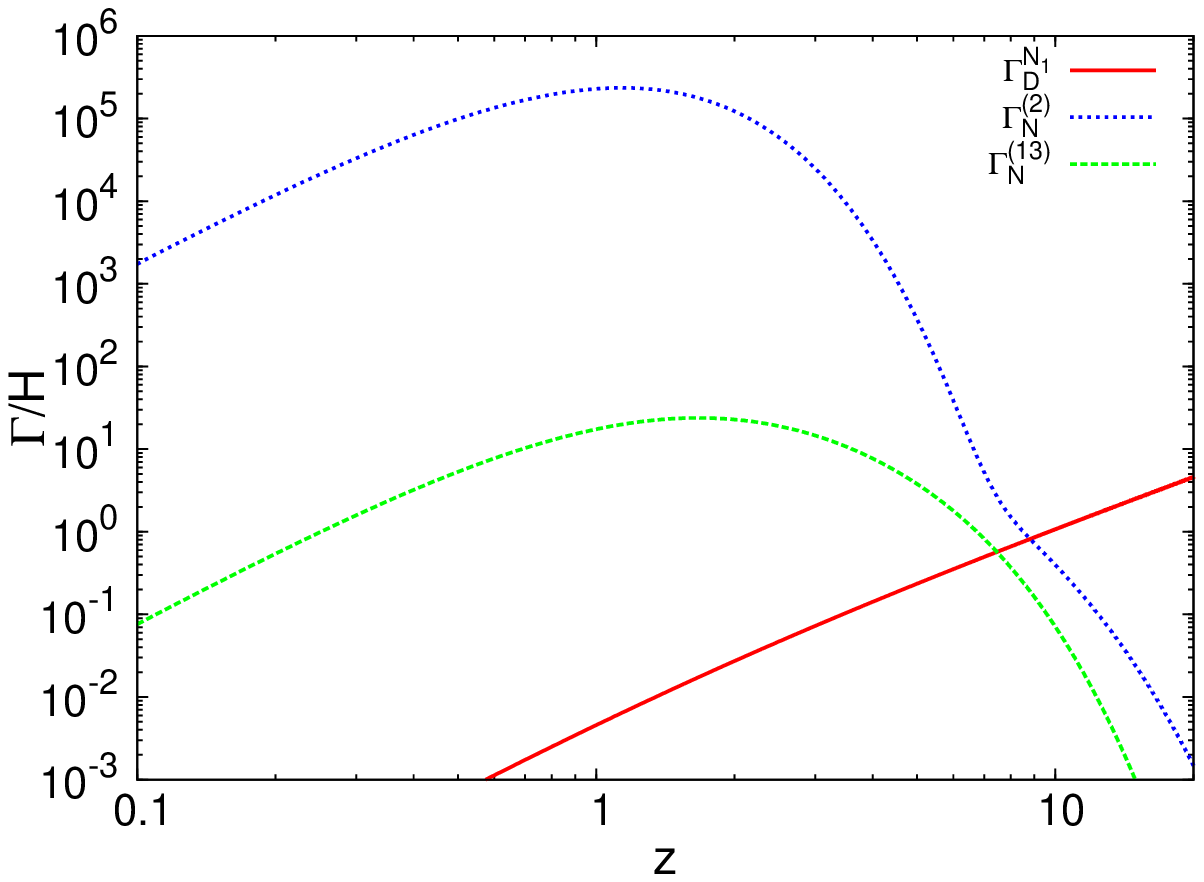}
\hspace*{5mm}
\epsfxsize=4.7cm
\leavevmode
\epsfbox{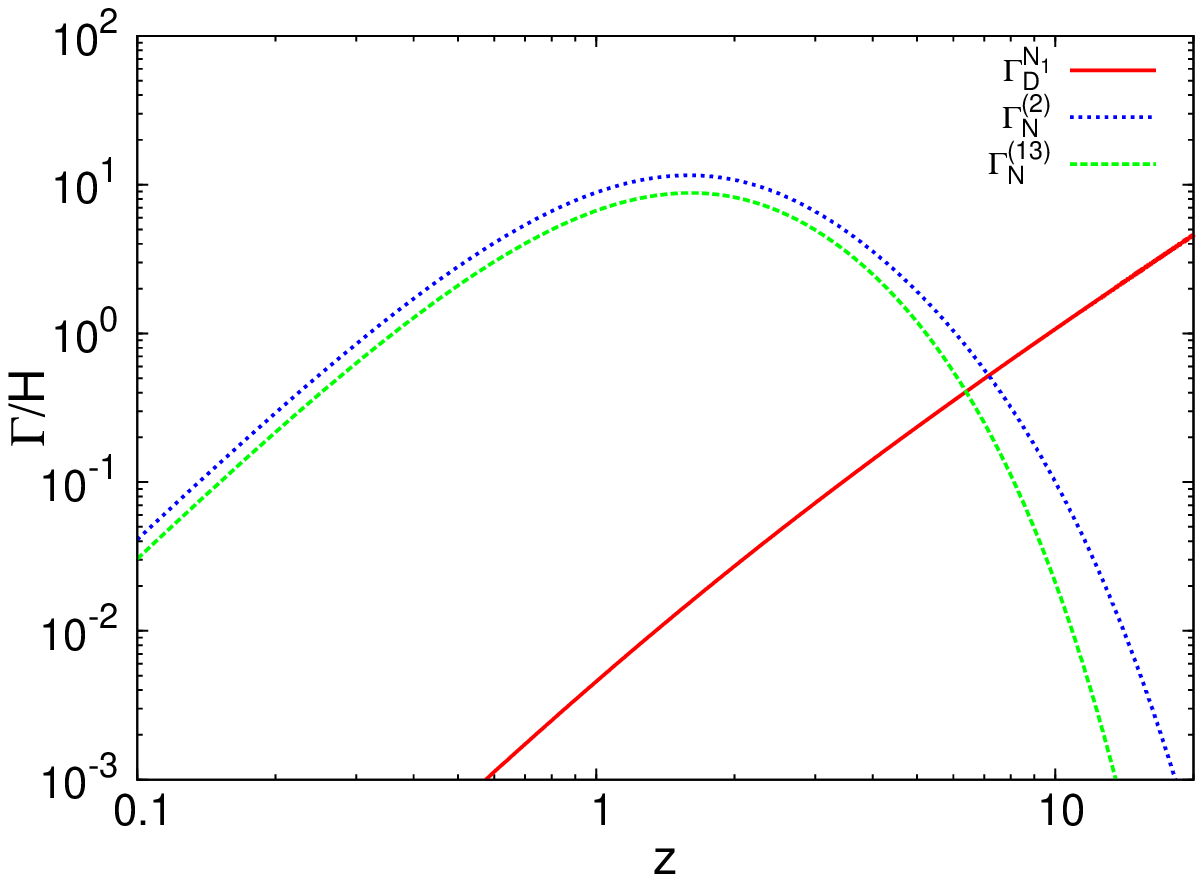}
\hspace*{5mm}
\epsfxsize=4.7cm
\leavevmode
\epsfbox{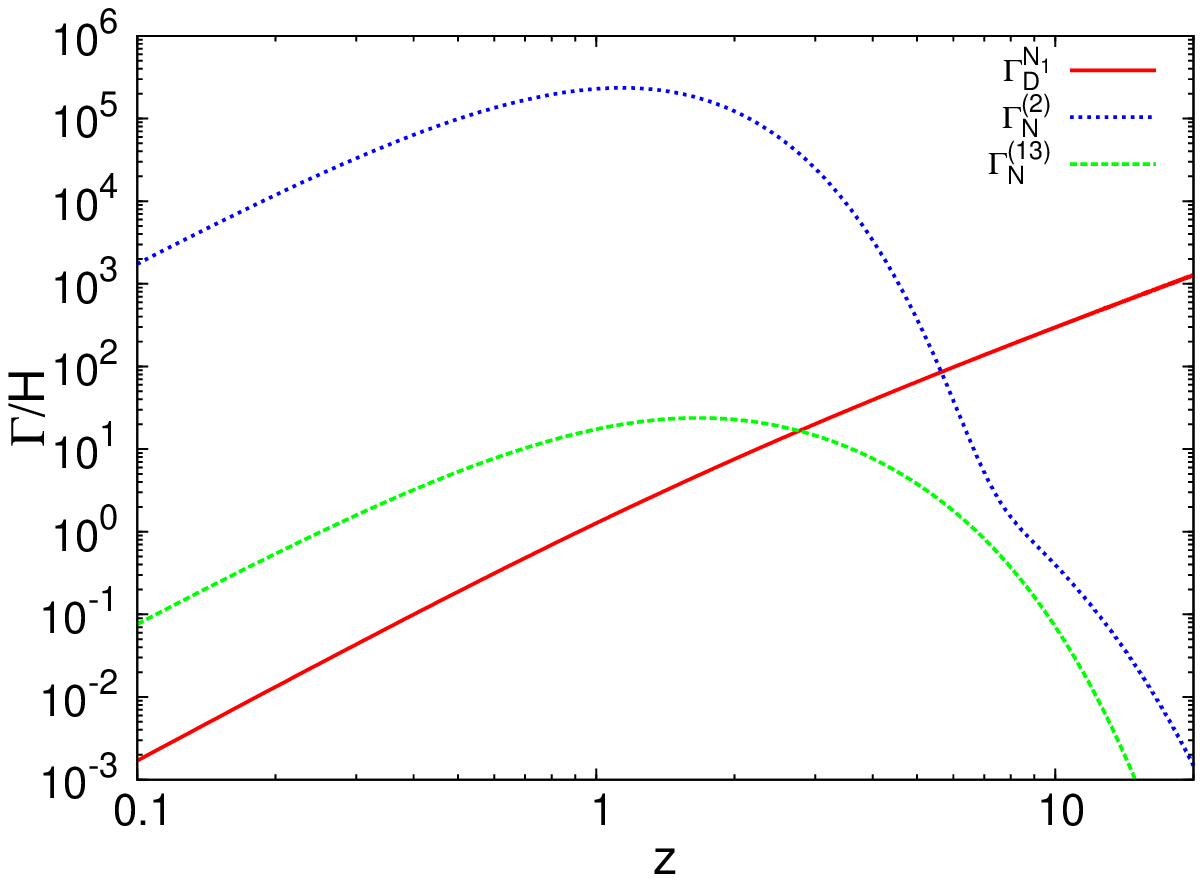} \\
\end{center}
\vspace*{-3mm}
{\footnotesize {\bf Fig.~3}~~The upper panels show the 
evolution of $Y_L$, $Y_{N_1}$ and 
$\Delta_{N_1}\equiv|Y_{N_1}-Y_{N_1}^{\rm eq}|$. 
The lower panels show the reaction rates 
$\frac{\gamma}{H}$ of the processes that have crucial effects for 
the leptogenesis in this model. We use the parameters
shown as (ia) and (ib) in Table 1 for the left and middle panels.
In the right panel the same parameters as (ia) are used except for
$|h_1|$, which is fixed to $|h_1|=5\cdot 10^{-7}$ in this case.
The black dotted line represents the required value for $|Y_L|.$
}
 \end{figure}

The solutions of eq.~(\ref{bqn}) are shown in the upper panels
of the left and the middle columns in Fig.~3 for cases (ia) 
and (ib) in Table 1.  
The generated lepton number asymmetry $|Y_L|$ is smaller than the value 
required for the explanation of the baryon number asymmetry 
at least by one order of magnitude.\footnote{The parameters used in 
case (ia) are almost equivalent to the one which is presented as the 
promising one for the generation of the sufficient baryon number 
asymmetry in \cite{idm1}.}
In case (ia), the $CP$ asymmetry parameter $|\varepsilon|$ takes 
rather large value such as $1.1\cdot 10^{-7}$.
This suggests that the washout of the generated lepton number 
asymmetry is effective.
In the figure of the right column we use the same parameters as 
the one of case (ia) except for
$|h_1|$ which is fixed to the larger value $5\cdot 10^{-7}$.
From this figure, we can see the role of this coupling which is
discussed above. 
Although this change does not affect the values of $|h_{2,3}|$ which explain
the neutrino oscillation data, we expect that
the deviation of the number density of $N_1$ from the equilibrium value 
becomes smaller than other cases with the smaller value of $|h_1|$. 
This is shown in the figures.
 
In the lower panel we plot the behavior of the relevant reaction rates 
for each case. 
These processes are crucial for the leptogenesis in this model.
The figures show that the lepton number-violating scatterings induced 
by the $s$-channel $N_i$ exchange are kept in the thermal equilibrium
until rather late period and the large part of the generated lepton
number asymmetry is washed out. 
This situation is common for all cases in Table 1. 
We give the predicted value of $Y_B$ for each case in the last 
column of Table~1. These examples show that the sufficient amount of baryon 
number asymmetry seems difficult to be generated through the thermal
leptogenesis in the present neutrino mass generation scheme at 
least as long as we impose the full neutrino oscillation data and the DM
direct search constraint.

\begin{figure}[t]
\begin{center}
\epsfxsize=7.5cm
\leavevmode
\epsfbox{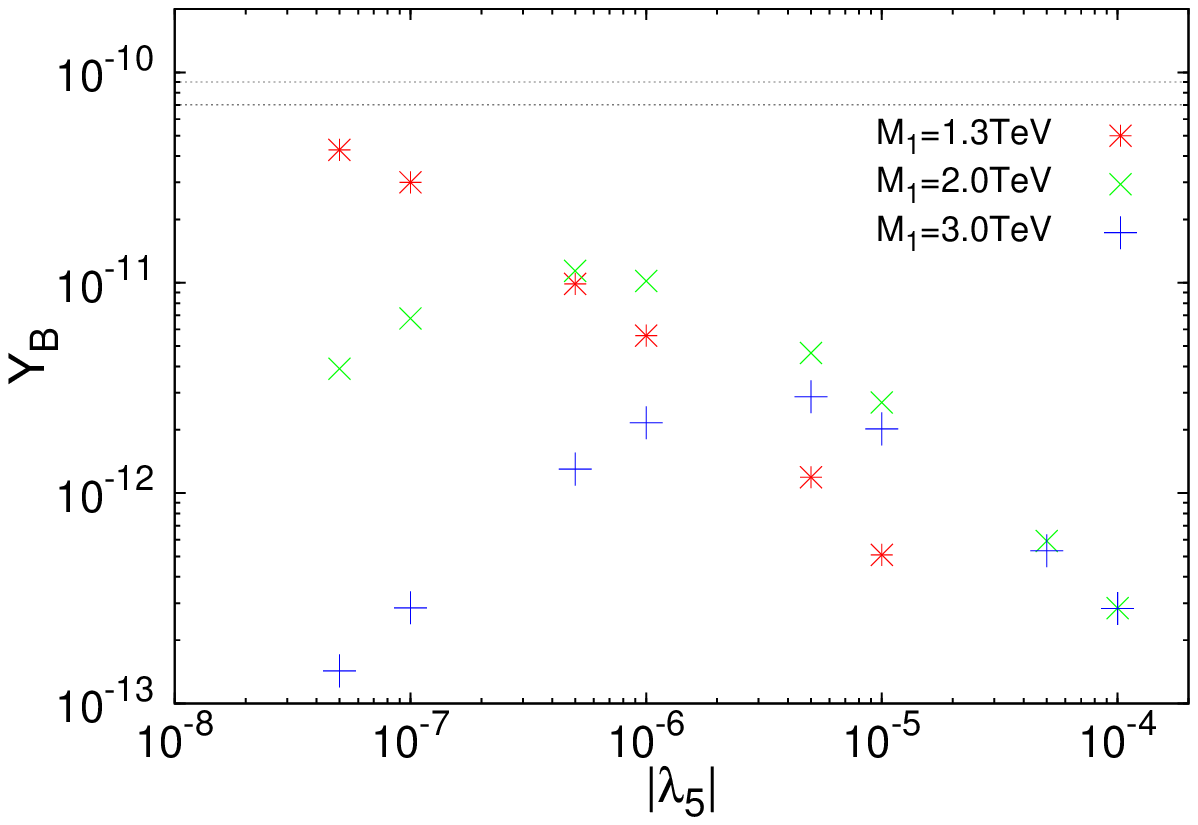}
\end{center}
\vspace*{-3mm}
{\footnotesize {\bf Fig.~4}~~The dependence of the generated 
baryon number asymmetry $Y_B$ on $|\lambda_5|$ in case (ia) with the
 different value of $M_1$. The value of $M_1$ is fixed to 1.3 TeV, 2 TeV
 and 3 TeV in each case.}
 \end{figure}

In order to confirm this statement in case (ia), we plot the 
generated baryon number asymmetry $Y_B$ for various values of 
$|\lambda_5|$ in Fig.~4.
The maximum value of $Y_B$ is found to be realized at a certain value of
$|\lambda_5|$ and it moves to the smaller $|\lambda_5|$
region for the smaller $M_1$.
This may be explained as follows. At the region with a larger
$|\lambda_5|$ value, the neutrino Yukawa couplings have smaller values to
give a smaller value for $|\varepsilon|$. On the other hand, at the 
region with a smaller $|\lambda_5|$ value, the neutrino Yukawa couplings 
have larger values to bring about the large washout.   
For the $M_1=1.3$~TeV case, the requited $Y_B$ can be obtained 
at a rather small value such as $|\lambda_5|\le 10^{-6}$ due to this nature.
However, it is excluded by the direct search
experiments as shown in eq.~(\ref{direct}).\footnote{It is useful to 
note that the smaller $M_\eta$ brings about the
smaller value for the lower bound of $|\lambda_5|$. However, its effect
is only a change of the factor in case of the high-mass $\eta$.}
Although we do not search the whole parameter space, 
we could say that the above-mentioned result does not change so easily 
for the values of $|\lambda_5|$ which satisfy the condition
(\ref{direct}). 
This is suggested by the fact that 
the value of $Y_B$ becomes smaller for the larger $|\lambda_5|$ 
which makes the neutrino Yukawa couplings smaller under the 
constraints of the neutrino oscillation data. 

Finally, we give some comments on the leptogenesis in the case where
the right-handed neutrinos have large masses comparable to the ones in 
the ordinary seesaw case.\footnote{This is considered in \cite{idm1a}. 
However, the neutrino oscillation data are not imposed 
in a quantitative way there.} 
In this case, we find that the DM relic abundance and the 
neutrino masses and mixing could be explained consistently by
setting $|\lambda_5|$ and the neutrino Yukawa couplings appropriately. 
The right-handed neutrino masses could be smaller than the ones 
in the ordinary tree-level seesaw case for the same neutrino Yukawa 
couplings since the neutrino masses are generated through the 
one-loop effect.
In the ordinary type I seesaw scenario with the hierarchical
right-handed neutrino masses, there is an upper bound
for the $CP$ asymmetry which is known as the Davidson-Ibarra (DI) 
bound \cite{di} and may be written as 
$|\varepsilon_{\rm DI}|=\frac{3}{8\pi}
\frac{M_1\sqrt{\Delta m^2_{\rm atm}}}{\langle\phi\rangle^2}$.
In the present case with the assumed flavour structure, 
the $CP$ asymmetry $|\varepsilon|$ is related to the DI bound as
$|\varepsilon|=|\varepsilon_{\rm DI}|\frac{\pi^2}{3\ln(M_2/M_\eta)}
\frac{1+q_1^2}{|\lambda_5|}\left(\frac{M_2}{M_1}\right)^2$.
This shows that the DI bound could be evaded depending on the value of
$|\lambda_5|$. However, this feature does not 
mean that the model causes
more efficient leptogenesis than the ordinary seesaw model.
Since the smaller $|\lambda_5|$ requires the larger neutrino Yukawa
couplings under the constraints of the neutrino oscillation data, 
the washout of the lepton number asymmetry is expected to become large. 

We examine this aspect under the assumed lepton flavor structure 
by imposing all the neutrino oscillation data quantitatively.  
We assume the hierarchical right-handed neutrino mass
spectrum and fix the parameters as follows,
\begin{equation}
\quad |h_1|= 10^{-4}, \quad M_1= 10^\alpha~{\rm GeV}, \quad 
 M_2= 10^{\alpha+1}~{\rm GeV}, \quad M_3=10^{\alpha+2}~{\rm GeV},
\end{equation}
where $|h_1|$ is determined by taking account of the condition (\ref{h1}).
We find that the $CP$ asymmetry $|\varepsilon|$ can be written for these
parameters as $|\varepsilon|\simeq|\varepsilon_{\rm DI}|\frac{2.5\cdot
10^2}{(\alpha-2)|\lambda_5|}$ where the DI bound $|\varepsilon_{\rm DI}|$ 
is given as $|\varepsilon_{\rm DI}|\le 1.9\cdot 
10^{-16+\alpha}$. 
This relation shows that the $CP$ asymmetry could escape the DI bound 
without causing the contradiction with the neutrino oscillation data if
$|\lambda_5|$ and $M_1$ take suitable values.
In order to fix the values of neutrino Yukawa couplings,
we impose the neutrino oscillation data in the same way as in the previous 
examples with $q_1=0.85$, $q_2=-0.27$, and $q_3=2.4$.
For such neutrino Yukawa couplings, we obtain $\sin^22\theta_{13}=0.085$ 
independently on the value of $|\lambda_5|$.  
The baryon number asymmetry $Y_B$ obtained at $z=20$ through the analysis of 
the Boltzmann equations is plotted for some typical values
of $|\lambda_5|$ and $M_1$ in the left panel of Fig.~5.
Since this $z$ is much smaller than $z_{\rm EW}=\frac{M_1}{T_{\rm EW}}$,
$Y_B(\infty)$ could be much smaller than the plotted value if the
washout effects do not decouple. 
However, we can confirm that the plotted $Y_B$ is recognized as
$Y_B(\infty)$, at least for $|\lambda_5|>|\lambda_5^{\rm max}|$, where
$|\lambda_5^{\rm max}|$ gives the maximum value of $Y_B$. 
In the region of $|\lambda_5|<|\lambda_5^{\rm max}|$, the neutrino
Yukawa couplings become large enough to continue reducing the generated lepton
number asymmetry through the lepton number-violating scatterings
which do not decouple at $z=20$ completely.  
Since $|h_{2,3}|$ is required to be larger for the smaller values of  
$|\lambda_5|$ from the neutrino oscillation data, 
the large part of the generated lepton number asymmetry 
is considered to be washed out effectively although a larger
$|\varepsilon|$ value is expected. 
The figure shows that the required value of $Y_B$ is generated for
$M_1>10^8$~GeV, which is somewhat smaller than the one required 
in the ordinary seesaw case.\footnote{The required baryon number
asymmetry could be generated even in the case with $M_1=O(10^8)$~GeV 
if special texture for the neutrino mass matrix is assumed even in the
ordinary seesaw case \cite{bs}.} 

\begin{figure}[t]
\begin{center}
\epsfxsize=7.5cm
\leavevmode
\epsfbox{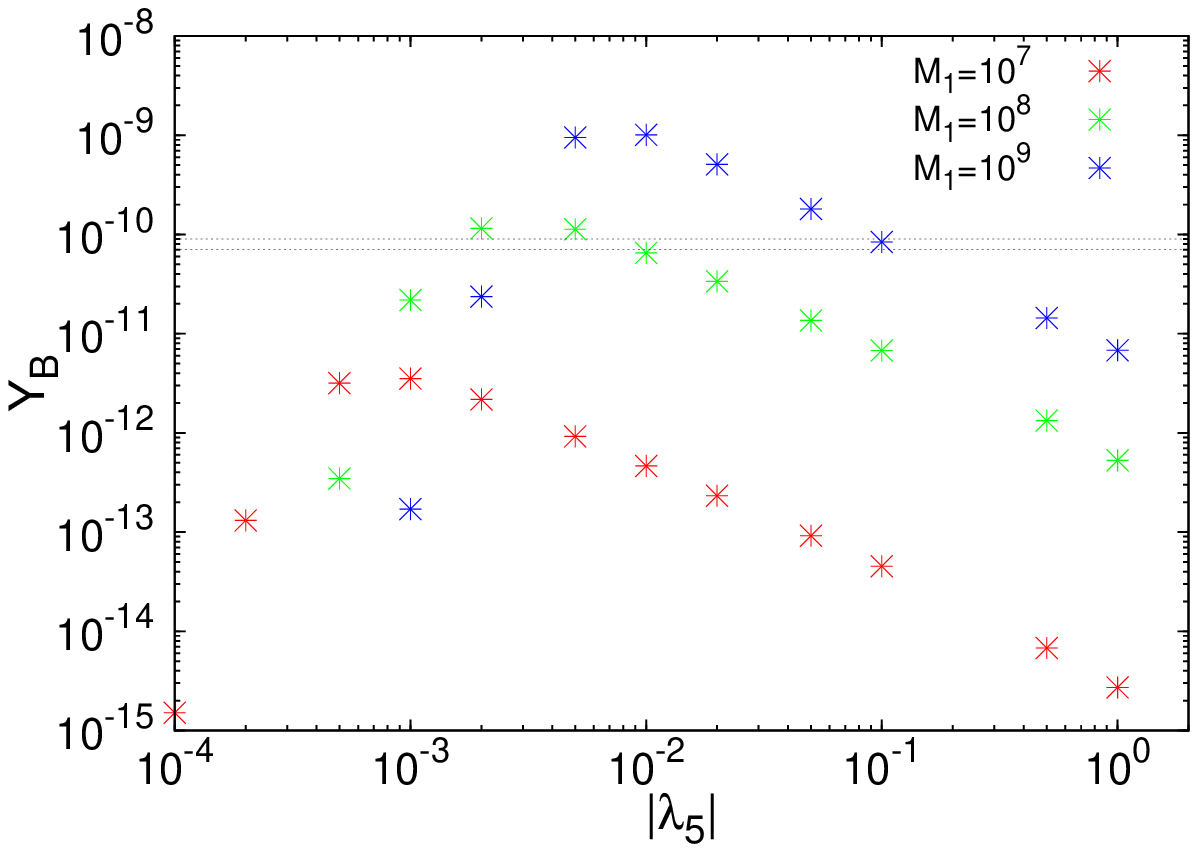}
\hspace*{4mm}
\epsfxsize=7.3cm
\leavevmode
\epsfbox{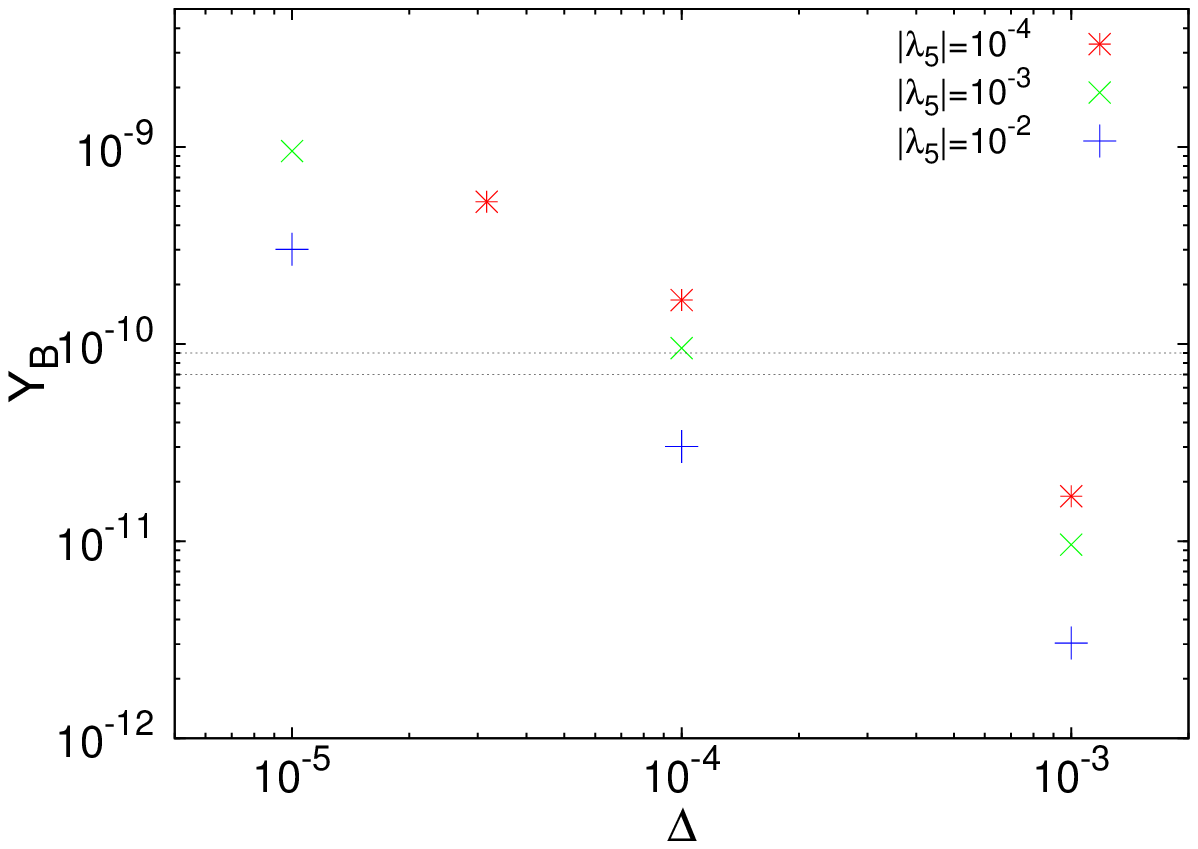}
\end{center}
\vspace*{-3mm}
{\footnotesize {\bf Fig.~5}~~The baryon number asymmetry $Y_B$ 
in cases of the heavy right-handed neutrinos (the left panel) and the 
degenerate right-handed neutrinos (the right panel). 
In the left panel, a GeV unit is used for the mass scale. In the right panel,
 $M_1=2$~TeV is assumed.}
 \end{figure}

\subsection{Improvement by suppressing the washout}
In the previous part we found that it is difficult to generate
the sufficient baryon number asymmetry in consistent with all the neutrino
oscillation data and the DM direct search as long as the mass of the 
lightest right-handed neutrino is assumed in a TeV range.
This result is considered to be brought about by the large washout 
effect of the generated lepton number asymmetry.   
Here, we consider a possible improvement of this situation 
by making the neutrino Yukawa couplings small enough to suppress the washout. 
In this improvement, the $CP$ asymmetry $|\varepsilon|$ should be kept to 
a suitable value such as $O(10^{-7})$ or more, simultaneously.  
Resonant leptogenesis can realize it.

As such an example, we suppose that the right-handed neutrino masses 
$M_1$ and $M_2$ are nearly degenerate in case (ia). $M_1$ is fixed to
2~TeV and $M_2$ is replaced with $M_2=(1+\Delta)M_1$. 
In this case, we can make the neutrino Yukawa couplings much smaller than 
the ones in case (ia) by assuming a larger value for $|\lambda_5|$.
For instance, if we put $|\lambda_5|=10^{-3}$, all the neutrino
oscillation data can be satisfied for $|h_2|=3.1\cdot 10^{-4}$ and
$|h_3|=1.5\cdot 10^{-4}$. 
Although the smaller neutrino Yukawa couplings tend to make the
$CP$ asymmetry $|\varepsilon|$ smaller, we can enhance $|\varepsilon|$ 
by supposing the degenerate right-handed neutrino masses.
On the other hand, in that case the washout effect could be suppressed 
sufficiently.
In the usual resonant leptogenesis with the right-handed neutrino 
masses at TeV regions \cite{epsilon,resonant,resonant1}, the generation 
of the sufficient lepton number asymmetry requires rather 
strict mass degeneracy. 
We examine whether this situation can be changed in this neutrino mass
generation scenario.

In the nearly degenerate right-handed neutrino case, the dominant 
contribution to the $CP$ asymmetry in the decay of $N_1$ comes 
from the interference between the tree and the self-energy 
diagrams.\footnote{Since the Yukawa couplings of $N_{2,3}$ 
are required to be much larger than $|h_1|=3\cdot 10^{-8}$ for $N_1$,  
$N_{2,3}$ are considered to be in the thermal equilibrium.  
Thus, we can not expect a substantial contribution to the lepton number 
asymmetry from the decays of $N_{2,3}$.} 
In the present case, the $CP$ asymmetry $|\varepsilon|$ can be expressed 
as \cite{epsilon,resonant,resonant1}
\begin{eqnarray}
\varepsilon&=&\sum_{i=2,3}\frac{{\rm Im}
(h^\dagger h)_{1i}^2}{(h^\dagger h)_{11}(h^\dagger h)_{ii}}
\frac{(M_1^2-M_i^2)M_1\Gamma_i}{(M_1^2-M_i^2)^2+M_1^2\Gamma_i^2} \nonumber\\
&\simeq&
\frac{(M_1^2-M_2^2)M_1\Gamma_2}{(M_1^2-M_2^2)^2+M_1^2\Gamma_2^2}
\sin 2(\varphi_1-\varphi_2)
\simeq-\frac{2\Delta\tilde\Gamma_2}{4\Delta^2+\tilde\Gamma_2^2}
\sin 2(\varphi_1-\varphi_2),
\end{eqnarray}
where $\tilde\Gamma_2=\frac{|h_2|^2}{4\pi}(1+q_1^2)
\left(1-\frac{M_\eta^2}{M_1^2}\right)^2$.
For case (ia) with $\sin 2(\varphi_1-\varphi_2)=O(1)$,
$|\varepsilon|$ can have a value of $O(1)$ for 
$\Delta\sim 5\cdot 10^{-7}$. Thus, we can expect that the 
mass degeneracy required to bring about the sufficient baryon number
asymmetry could be much milder than the usually assumed 
value $\Delta~{^<_\sim}~10^{-8}$ in the TeV-scale resonant 
leptogenesis \cite{resonant1}.

We use the formula for $\varepsilon$ given above and estimate $Y_B$
by solving the Boltzmann equations in eq.~(\ref{bqn})
for some typical values of $\Delta$ by varying the value of $|\lambda_5|$.
In the right panel of Fig.~5, we plot the numerical results of the 
obtained baryon number asymmetry.  
The required baryon number asymmetry is found to be generated 
for the right-handed neutrinos with the mass degeneracy 
$\Delta=O(10^{-4})$ for each value of $|\lambda_5|$. 
This degeneracy is much milder in comparison with the ordinary resonant 
leptogenesis at TeV scales. 
The result is caused by the nature of the model such that 
the neutrino Yukawa couplings can take sufficiently small values 
to suppress the washout effect keeping the neutrino masses 
in the appropriate range for the explanation of the neutrino oscillation data.
The freedom of $\lambda_5$ makes it possible. 
The present neutrino mass generation scheme can give the consistent 
explanation for the three phenomenological problems in the SM if 
only the rather mild mass degeneracy between two light right-handed neutrinos
is assumed without large extension of the model.
  
\section{Summary}
The inert doublet model extended with the right-handed neutrinos 
is a simple and 
interesting framework for both neutrino masses and dark matter.
In this scenario, the lightest right-handed neutrino or 
the lightest neutral component of the
inert doublet can be a dark matter candidate. Since the neutrino Yukawa
couplings should be $O(1)$ to reduce the relic abundance of dark matter in the
former case, the thermal leptogenesis is difficult due to the strong
washout effect. On the other hand, the neutrino Yukawa couplings 
can be irrelevant to the dark matter abundance in the latter case. 
Thus, the latter scenario has been considered to give the consistent 
explanation for the origin of the small neutrino masses, the existence
of dark matter and the baryon number asymmetry in the Universe.

In this paper, we reexamined the possibility for the simultaneous 
explanation of these three problems in the latter case. 
We took account of the quantitative explanation for 
all the neutrino oscillation data including the recent results 
for $\sin^22\theta_{13}$ in the analysis.  
The results of our study suggest 
that the sufficient amount of baryon number asymmetry seems
not to be generated in a consistent way with the full 
neutrino oscillation data. 
The neutrino Yukawa couplings could be large enough to enhance the 
$CP$ asymmetry even for the $O(1)$~TeV right-handed neutrinos 
in the consistent way with the neutrino oscillation 
data by using the freedom of $\lambda_5$.  However, the same Yukawa
couplings induce the lepton number-violating scattering processes to
wash out the generated lepton number asymmetry. 
We checked this point through the numerical study. 
Although we do not study the whole parameter space, this feature seems
to be rather general and then it seems not so easy to find the parameters 
to escape this situation without any modification.  

We also examined the same problem in the modified situation such as
the case with the heavy right-handed neutrinos like the ordinary type I
seesaw and also the case with the degenerate
light right-handed neutrinos.
In the former case, we found that the right-handed neutrino 
could be somewhat lighter than the ordinary seesaw one but 
needs to be heavy enough to the similar level to it.
In the latter case, even if the right-handed neutrino masses are in a
TeV range, the resonant effect can enhance the $CP$
asymmetry to generate the sufficient amount of baryon number asymmetry
even for the small neutrino Yukawa couplings which can suppress the
washout sufficiently. 
An interesting point is that the required mass degeneracy in this case 
is much milder than the one in the ordinary seesaw case. 
This possibility is brought about
by the characteristic feature in this inert doublet model with 
the radiative neutrino mass generation.
More complete study of the parameter regions which are not searched 
in this paper will be presented in future publication. 
   
\section*{Acknowledgement}
This work is partially supported by Grant-in-Aid for Scientific
Research (C) from Japan Society for Promotion of Science (No. 21540262
and No. 24540263), and also a Grant-in-Aid for Scientific Research 
on Priority Areas from The Ministry of Education, Culture, Sports, 
Science and Technology (No. 22011003).

\newpage
\section*{Appendix}
In this Appendix, we give the formulas of the reaction density
contributing to the Boltzmann equations for the number density 
of $N_1$ and the lepton number asymmetry.
For the processes relevant to their evolution, we could refer to 
the reaction density given in \cite{cross}.
In the present model, however, interaction terms of $\eta$ and $N_1$ 
are restricted by the $Z_2$ symmetry and also the masses of 
$\eta$ and $N_i$ take the similar order values, which cause large difference 
from the ordinary seesaw leptogenesis.  
Thus, we need to modify these formulas by taking account of the features 
of the present model.\footnote{Although the modified ones are
given in Appendix of \cite{ndm1}, the mass spectrum assumed there is
different from the present one. The following formulas are arranged to
applicable to the scenario in this paper. }

In order to give the expression for the reaction density of the relevant
processes, we introduce dimensionless variables
\begin{equation}
x=\frac{s}{M_1^2}, \qquad a_j=\frac{M_j^2}{M_1^2}, \qquad 
a_\eta=\frac{M_\eta^2}{M_1^2},
\end{equation}
where $s$ is the squared center of mass energy.
The reaction density for the decay of $N_1$ can be 
expressed as
\begin{equation}
\gamma_D^{N_1}=\frac{(1+q_1^2)|h_1|^2}{4\pi^3}
M_1^4(1-a_\eta)^2\frac{K_1(z)}{z}, 
\label{decay}
\end{equation} 
where $K_1(z)$ is the modified Bessel function of the first kind.

The reaction density for the scattering processes is expressed as
\begin{equation}
\gamma(ab\rightarrow ij)=\frac{T}{64\pi^4}\int^\infty_{s_{\rm min}}ds~
\hat\sigma(s)\sqrt{s}K_1\left(\frac{\sqrt{s}}{T}\right),
\end{equation}
where
$s_{\rm min}={\rm max}[(m_a+m_b)^2,(m_i+m_j)^2]$ and 
$\hat\sigma(s)$ is the reduced cross section. 
In order to give the expression for the reaction density of the
processes relevant to eq.~(\ref{bqn}), we define the following 
quantities for convenience:
\begin{eqnarray}
&& \frac{1}{D_i(x)}=\frac{x-a_i}{(x-a_i)^2+a_i^2c_i}, \qquad 
c_i=\frac{1}{16\pi^2}\left(\sum_{k=e,\mu,\tau}
|h_{ki}|^2\right)^2\left(1-\frac{a_\eta}{a_i}\right)^4, \nonumber \\
&&\lambda_{ij}=\left[x-(\sqrt{a_i}+\sqrt{a_j})^2\right]
\left[x-(\sqrt{a_i}-\sqrt{a_j})^2\right],
 \nonumber \\
&&L_{ij}=\ln\left[\frac{x-a_i-a_j+ 2a_\eta +\sqrt{\lambda_{ij}}}
{x-a_i-a_j +2 a_\eta -\sqrt{\lambda_{ij}}}\right], \nonumber \\
&&L_{ij}^\prime=\ln\left[\frac{\sqrt{x}(x-a_i-a_j-2a_\eta)
+\sqrt{\lambda_{ij}(x-4a_\eta)}}
{\sqrt{x}(x-a_i-a_j-2a_\eta) 
-\sqrt{\lambda_{ij}(x-4a_\eta)}}\right].
\end{eqnarray}

As the lepton number violating scattering processes induced through the
$N_i$ exchange,
we have
\begin{eqnarray}
\hat\sigma^{(2)}_N(x)&=&\frac{1}{2\pi}\frac{(x-a_\eta)^2}{x^2}
\left[\sum_{i=1}^3(hh^\dagger)_{ii}^2\frac{a_i}{x}
\left\{\frac{x^2}{xa_i -a_\eta^2}+\frac{2x}{D_i(x)}
+\frac{(x-a_\eta)^2}{2D_i(x)^2}\right.\right.\nonumber \\
&-&\left.\frac{x^2}{(x-a_\eta)^2}
\left(1+\frac{2(x+a_i)-4a_\eta}{D_i(x)}\right)
\ln\left(\frac{x(x+a_i-2a_\eta)}{xa_i-a_\eta^2}\right)\right\}\nonumber \\
&+&\left.
\sum_{i>j}{\rm Re}[(hh^\dagger)_{ij}^2]\frac{\sqrt{a_ia_j}}{x}\left\{
\frac{x}{x-a_i}+\frac{x}{x-a_j}+\frac{(x-a_\eta)^2}{(x-a_i)(x-a_j)}
\right.\right.\nonumber \\
&+&\left.\left.\frac{x^2}{(x-a_\eta)^2}
\left(\frac{2(x+a_i-2a_\eta)}{a_j-a_i}-
\frac{x+a_i-2a_\eta}{x-a_j}\right)\ln\frac{x(x+a_i-2a_\eta)}{xa_i-a_\eta^2}
\right.\right. \nonumber\\
&+&\left.\left.\frac{x^2}{(x-a_\eta)^2}
\left(\frac{2(x+a_j-2a_\eta)}{a_i-a_j}-
\frac{x+a_j-2a_\eta}{x-a_i}\right)\ln\frac{x(x+a_j-2a_\eta)}{xa_j-a_\eta^2}
\right\}\right]
\label{lv1}
\end{eqnarray}
for $\ell_\alpha\eta^\dagger \rightarrow \bar\ell_\beta\eta$ and also
\begin{eqnarray}
\hat\sigma^{(13)}_N(x)&=&\frac{1}{2\pi}
\left[\sum_{i=1}^3(hh^\dagger)^2_{ii}\left\{
\frac{a_2(x^2-4xa_\eta)^{1/2}}{a_ix+(a_i-a_\eta)^2}\right.\right. \nonumber \\
&+&\left.\left.
\frac{a_i}{x+2a_i-2a_\eta}
\ln\left(\frac{x+(x^2-4xa_\eta)^{1/2}+2a_i-2a_\eta}
{x-(x^2-4xa_\eta)^{1/2}+2a_i-2a_\eta}\right)\right\}
\right.\nonumber\\
&+&\left.
\sum_{i>j}
\frac{{\rm Re}[(hh^\dagger)_{ij}^2]\sqrt{a_ia_j}}{x+a_i+a_j-2a_\eta}
\ln\left(\frac{x+(x^2-4xa_\eta)^{1/2}+a_i+a_j-2a_\eta}
{x-(x^2-4xa_\eta)^{1/2}+a_i+a_j-2a_\eta}\right)\right] 
\label{lv2}
\end{eqnarray}
for $\ell_\alpha\ell_\beta \rightarrow \eta\eta$.
Here we note that cross terms has no contribution if the maximum $CP$
phases are assumed in the way as $\sin2(\varphi_{2,3}-\varphi_1)=1$ 
with $\varphi_1=0$. We adopt this possibility in the numerical analysis,
for simplicity. 
Since another type of lepton number violating processes 
$N_{i}N_{j}\rightarrow \ell_\alpha\ell_\beta$ induced 
by the $\eta$ exchange have additional suppression due to a small 
$|\lambda_5|$, we can neglect them safely.\footnote{We should note that
$|\lambda_5|$ might not have a small value in the case with heavy
right-handed neutrinos, which is discussed in section 3.
In this case, these processes could give large contribution to the
washout of the generated lepton number asymmetry.}
 
As the lepton number conserving scattering processes which contribute to
determine the number density of $N_1$, we have
\begin{eqnarray}
\hat\sigma^{(2)}_{N_iN_j}(x)&=&\frac{1}{4\pi}
\left[\sum_{i,j=1}^3|(hh^\dagger)_{ii}(hh^\dagger)_{jj}
\frac{\sqrt{\lambda_{ij}}}{x}\left(1+
\frac{(a_i-a_\eta)(a_j-a_\eta)}{(a_i-a_\eta)(a_j-a_\eta)+xa_\eta}
\right.\right.\nonumber \\
&+&\left.\left.\frac{a_i+a_j-2a_\eta}{x}L_{ij}\right)
- {\rm Re}[(h h^\dagger)_{ij}^2]
\frac{2\sqrt{a_ia_j}L_{ij}}{x-a_i-a_j+2a_\eta}\right]
\label{nlv1}
\end{eqnarray}
for $N_{i}N_{j} \rightarrow \ell_\alpha\bar\ell_\beta$ which are 
induced through the $\eta$ exchange and also
\begin{eqnarray}
\hat\sigma^{(3)}_{N_iN_j}(x)&=&
\frac{1}{4\pi}\frac{(x-4a_\eta)^{1/2}}{x^{1/2}}
\left[|(hh^\dagger)_{ij}|^2\left\{
\frac{\sqrt{\lambda_{ij}}}{x}
\Big(-2  \right.\right. \nonumber \\
&+&\left.\frac{4a_\eta(a_i- a_j)^2}
{(a_\eta-a_i)(a_\eta-a_j)x +(a_i-a_j)^2a_\eta}\Big) 
+\left(1-2\frac{a_\eta}{x}\right)
L_{ij}^\prime\right\}  \nonumber\\
&-&\left. {\rm Re}[(h h^\dagger)_{ij}^2]
\left(\frac{\sqrt{\lambda_{ij}}}{x} +
\frac{2(a_\eta^2-a_ia_j)L_{ij}^\prime}
{(x^2-4xa_\eta)^{1/2}(x-a_i-a_j-2a_\eta)}
\right)\right]
\label{nlv2}
\end{eqnarray}
 for $N_iN_j \rightarrow \eta\eta^\dagger$ which are 
induced through the $\ell_\alpha$ exchange.
The cross terms in these reduced cross sections are neglected 
because of the same reasoning as eqs~(\ref{lv1}) and (\ref{lv2}). 

In order to see the behavior of these relevant processes such as the
decoupling time, we may estimate the ratio of the reaction rate to the Hubble
rate $\frac{\Gamma}{H}$ as a function of $z$ (see Fig.~3). 
The thermally averaged reaction rate $\Gamma$ is related to
the above discussed reaction densities through
\begin{equation}
\Gamma_D^{N_1}=\frac{\gamma_D^{N_1}}{n_{N_{1}}^{\rm eq}}
\end{equation}
for the decay of $N_{1}$ and also
\begin{equation}
\Gamma_N^{(2,13)}=\frac{\gamma_N^{(2,13)}}{n_\ell^{\rm eq}}, \qquad  
\Gamma_{N_iN_1}^{(2,3)}=\frac{\gamma_{N_iN_1}^{(2,3)}}{n_{N_{R_1}}^{\rm eq}}
\end{equation}
for the 2-2 scattering processes given in eqs.~(\ref{lv1}), (\ref{lv2})
and eqs.~(\ref{nlv1}), (\ref{nlv2}), respectively.  

\newpage
\bibliographystyle{unsrt}

\begin{thebibliography}{99}
\bibitem{nexp}Super-Kamiokande Collaboration, Y.~Fukuda, {\it et al.},
	Phys. Rev. Lett. {\bf 81} (1998) 1562; 
       SNO Collaboration, Q.~R~.Ahmad, {\it et al.},
	Phys. Rev. Lett. {\bf 89} (2002) 011301;
         KamLAND Collaboration, K.~Eguchi, {\it et al.}, 
       Phys. Rev. Lett. {\bf 90} (2003)
	021802; 
       K2K Collaboration, M.~H.~Ahn, {\it et al.},
	Phys. Rev. Lett. {\bf 90} (2003) 041801.    

\bibitem{uobs}WMAP Collaboration, D.~N.~Spergel, {\it et al.},
	Astrophys. J. {\bf 148} (2003) 175; 
SDSS Collaboration, M.~Tegmark, {\it et al.}, 
Phys. Rev. {\bf D69} (2004) 103501.

\bibitem{basym}A.~Riotto and M.~Trodden,
	Ann. Rev. Nucl. Part. Sci. {\bf 49} (1999) 35; 
W.~Bernreuther, Lect. Notes
	Phys. {\bf 591} (2002) 237; M.~Dine and A.~Kusenko,
	Rev. Mod. Phys. {\bf 76} (2003) 1.

\bibitem{ma}E.~Ma, Phys. Lett. {\bf B625} (2005) 76. 

\bibitem{idm}R.~Barbieri, L.~J.~Hall and V.~S.~Rychkov, Phys. Rev. {\bf
	D74} (2006) 015007; M.~Cirelli, N.~Fornengo and A.~Strumia,
	Nucl. Phys. {\bf B753} (2006) 178; L.~L.~Honorez, E.~Nezri,
	J.~F.~Oliver and M.~H.~G.~Tytgat, JCAP {\bf 0702} (2007) 028; 
Q.-H.~Cao and E.~Ma, Phys. Rev. {\bf D76} (2007) 095011;
S.~Andreas, M.~H.~G.~Tytgat and Q.~Swillens, JCAP {\bf 0904} (2009) 004; 
E.~Nezri, M.~H.~G.~Tytgat and G.~Vertongen, JCAP {\bf 0904} (2009) 014;
L.~L.~Honorez, JCAP {\bf 1101} (2011) 002; M.~Gustafsson, S.~Rydbeck,
	L.~L.~Honorez, and E.~Lundstr\"{o}m, arXiv:1206.6316 [hep-ph].

\bibitem{idm1}T.~Hambye, F.-S.~Ling, L.~L.~Honorez and J.~Roche, JHEP
	{\bf 0907} (2009) 090.  

\bibitem{idm1a}E.~Ma, Mod. Phys. Lett. {\bf A21} (2006) 1777.  

\bibitem{inel}D.~T.-Smith and N.Weiner, Phys.Rev.{\bf D72} (2005) 063509; 
S.~Chang, G.~D.~Kribs D.~T.-Smith and N.~Weiner, Phys. Rev. {\bf D79}
	(2009) 043513

\bibitem{l5}Y.~Cui, D.~E.~Marrissey, D.~Poland and L.~Randall, JHEP {\bf
	0905} (2009) 076;
C.~Arina, F.-S.~Ling and M.~H.~G.~Tytgat, JCAP {\bf 0910}
	(2009) 018.

\bibitem{ndm}E.~Ma, Phys. Rev. {\bf D73} (2006) 077301;
J.~Kubo, E.~Ma and D.~Suematsu, Phys. Lett. {\bf B642} (2006) 18;
J.~Kubo and D.~Suematsu, Phys. Lett. {\bf B643} (2006) 336;
D.~Aristizabal Sierra, J.~Kubo, D.~Restrepo,
D.~Suematsu and O.~Zapata, Phys. Rev. {\bf D79} (2009) 013011;
D.~Suematsu, T.~Toma and T.~Yoshida, Phys. Rev. {\bf D79} (2009) 093004; 
D.~Suematsu, T.~Toma and T.~Yoshida, Phys. Rev. {\bf D82} (2010) 013012. 

\bibitem{susyndm}H.~Fukuoka, J.~Kubo and D.~Suematsu, Phys. Lett. {\bf B678}
	(2009) 401; D.~Suematsu and T.~Toma, Nucl. Phys. {\bf B847}
	(2011) 567; H.~Fukuoka, D.~Suematsu and T.~Toma, 
JCAP {\bf 07} (2011) 001.

\bibitem{ndm1}D.~Suematsu, Eur. Phys. J {\bf C72} (2012) 72.

\bibitem{ndmext}D.~Suematsu, Eur. Phys. J.{\bf C56} (2008) 379;
 H.~Higashi, T.~Ishima and D.~Suematsu, Int. J. Mod. Phys. {\bf A26}
	(2011) 995; D.~Suematsu, Phys. Rev. {\bf D85} (2012) 073008.

\bibitem{t13}T2K Collaboration, K.~Abe, {\it et al.},
Phys. Rev. Lett. {\bf 107} (2011) 041801; 
Double Chooz Collaboration, Y.~Abe, {\it et al.},
	Phys. Rev. Lett. {\bf 108} (2012) 131801;
RENO Collaboration, J.~K.~Ahn, {\it et al.}, 
Phys. Rev. Lett. {\bf 108} (2012) 191802;
The Daya Bay Collaboration, F.~E.~An, {\it et al.},
	Phys. Rev. Lett. {\bf 108} (2012) 171803. 

\bibitem{relic}K.~Griest and D.~Seckel, Phys. Rev. {\bf D43} (1991)
	3191; P.~Gondolo and G.~Gelmini, Nucl. Phys. {\bf B360} (1991) 145.

\bibitem{inelvel}D.~Tucker-Smith and N.~Weiner, Phys. Rev. {\bf D64} (2001)
	043502.

\bibitem{direct1}CDMS Collaboration, Z.~Ahmed, {\it et al.},
	Phys. Rev. Lett. {\bf 102} (2009) 011301; XENON100
	Collaboration, E.~Aprile, {\it et al.}, Phys. Rev. Lett. {\bf
	105} (2010) 131302.

\bibitem{direct2}G.~Angloher {\it et al.}, Astropart. Phys. {\bf 31}
	(2009) 270; V.~N.~Lebedenko {\it et al.}, Phys. Rev. {\bf D80}
	(2009) 052010.

\bibitem{prg}K.~Nakamura {\it et al.} (Particle Data Group), 
J. Phys. G 37 (2010) 075021. 

\bibitem{bestf}T.~Schwetz, M.~T\'ortola and J.~Valle, New. J. Phys. {\bf
	10} (2008) 113011; arXiv:1103.0734.

\bibitem{fy}M.~Fukugita and T.~Yanagida, Phys. Lett. {\bf B174} (1986) 45.

\bibitem{epsilon}A.~Pilaftsis, Phys. ReV. {\bf D56} (1997) 5431.


\bibitem{kt}E.~W.~Kolb and S.~Wolfram, Nucl. Phys. {\bf B172} (1980)
	224; E.~W.~Kolb and M.~S.~Turner, {\it The Early Universe}
	(Addison-Wesley, Redwood City, CA, 1990).
	
\bibitem{cross}M.~Luty, Phys. Rev. {\bf D45} (1992) 455; M.~Plumacher,
Nucl. Phys. {\bf B530} (1998) 207.	
 
\bibitem{di}S.~Davidson and A.~Ibarra, Phys. Lett. {\bf B535} (2002) 25.

\bibitem{bs}T.~Baba and D.~Suematsu, Phys. Rev. {\bf D71} (2005) 073005.

\bibitem{resonant}M.~Flanz, E.~A.~Pascos and U.~Sarkar, Phys. Lett. {\bf
	B345} (1995) 248; L.~Covi, E.~Roulet and F.~Vissani,
	Phys. Lett. {\bf B384} (1996) 169; 
E.~Akhmedov, M.~Frigerio and A. Yu Smirnov, JHEP {\bf 0309} (2003) 021;
C.~H.~Albright and S.~M.~Barr, Phys. Rev. {\bf D69} (2004) 073010; 
T.~Hambye,J.~March-Russell and S.~W.~West, JHEP {\bf 0407} (2004) 070.

\bibitem{resonant1} A.~Pilaftsis and E.~J.~Underwood, 
Nucl. Phys. {\bf B692} (2004) 303; 
A.~Pilaftsis and E.~J.~Underwood, Phys. Rev. {\bf D72} (2005) 113001.

\end{thebibliography}

\end{document}